\documentclass[11pt,a4paper]{article}
\usepackage{amsmath,bm,graphicx,booktabs}
\usepackage{geometry, float,appendix,lmodern}
\usepackage{multirow}
\usepackage[authoryear]{natbib}
\usepackage{algorithm,algorithmic}
\usepackage[colorlinks,citecolor=blue,urlcolor=blue]{hyperref}
\usepackage{inputenc}
\usepackage[english]{babel}
\usepackage{amsthm}
\usepackage{color}
\usepackage{courier}
\usepackage{amssymb}
\usepackage{mathrsfs}
\usepackage{caption}
\usepackage{subcaption}
\usepackage{enumitem}
\usepackage{multirow}
\graphicspath{{figures/}}

\usepackage{authblk}

\providecommand{\keywords}[1]{\textbf{\textit{key words: }} #1}
\newcommand{\pcite}[1]{\citeauthor{#1}'s (\citeyear{#1})}

\newtheorem{theorem}{Theorem}

\newtheorem{remark}{Remark}
\newtheorem{corollary}{Corollary}

\DeclareMathOperator*{\argmin}{arg\,min}

\begin{document}

\title{Spatial Heterogeneity Automatic Detection and Estimation}
\author[1]{Xin Wang\thanks{Email: wangx172@miamioh.edu}}
\author[2]{Zhengyuan Zhu\thanks{Email: zhuz@iastate.edu}}
\author[3]{Hao Helen Zhang\thanks{Email:  hzhang@math.arizona.edu}}
\affil[1]{Department of Statistics, Miami University}
\affil[2]{Department of Statistics, Iowa State University}
\affil[3]{Department of Mathematics, University of Arizona}

\date{}
\maketitle
\begin{abstract}
Spatial regression is widely used for modeling the relationship between a dependent variable and explanatory covariates. Oftentimes, the linear relationships vary across space, when some covariates have location-specific effects on the response. One fundamental question is how to detect the systematic variation in the model and identify which locations share common regression coefficients and which do not. Only a correct model structure can assure unbiased estimation of coefficients and valid inferences. In this work, we propose a new procedure, called Spatial Heterogeneity Automatic Detection and Estimation (SHADE), for automatically and simultaneously subgrouping and estimating covariate effects for spatial regression models. The SHADE employs a class of spatially-weighted fusion type penalty on all pairs of observations, with location-specific weight adaptively constructed using spatial information, to cluster coefficients into subgroups. Under certain regularity conditions, the SHADE is shown to be able to identify the true model structure with probability approaching one and estimate regression coefficients consistently. We develop an alternating direction method of multiplier algorithm (ADMM) to compute the SHAD efficiently. In numerical studies, we demonstrate empirical performance  of the SHADE by using different choices of weights and compare their accuracy. The results suggest that spatial information can enhance subgroup structure analysis in challenging situations when the spatial variation among regression coefficients is small or the number of repeated measures is small. Finally, the SHADE is applied to find the relationship between a natural resource survey and a land cover data layer to identify spatially interpretable groups.
\end{abstract}


\section{Introduction}
\label{sec_intro}

Spatial regression is commonly used to model the relationship between a response and explanatory variables. For complex problems, some covariates (we call them global covariates) may have constant effects across space, while other covariates (we call them local covariates) may have location-specific effects, i.e, their effects on the response variable vary across space. This has received wide attention in many fields such as environmental sciences \citep{hu2018bayesian}, biology \citep{zhang2011bayesian}, social science \citep{bradley2018computationally}, economics \citep{brunsdon1996geographically}, and biostatistics \citep{xu2019latent}.  

A motivating example is about studying the relationship between two landcover data sources, one is the National Resources Inventory (NRI, \citealt{nusser1997national}) survey conducted by the USDA 
Natural Resources Conservation Service (NRCS), the other one is the Cropland data layer (CDL, \citealt{han2012cropscape}) produced by the USDA National Agricultural Statistics Service (NASS). Accurate estimate on local landcover information from NRI is essential for developing conservation policies and land management plans. However, direct estimates in small geographical areas such as at the county level may not be accurate due to small sample sizes. Auxiliary information such as CDL can be used to improve the small area estimator in NRI \citep{wang2018small}. Traditional regression models used in the small area estimation problems typically assume common regression coefficients over all domains, which may not be appropriate.  
For example, when we looked at the linear relationship between the NRI and CDL estimates of different types of land covers at county level, 
the regression coefficients in the Mountain states are quite different from the west coast and the vast areas in the east. This is reflected in Figure \ref{fig_mapest} (a) in Section \ref{sec_example}. One reason for that difference is due to the scope of the NRI survey, which only include non-federal land in the US. Another contributing reason is that CDL is created by training separate machine learning models at the state level using only ground observations from that state, which creates variations among states.
A common regression assumption would be too simple to capture the regional differences and will lead to biases in the estimators. This type of spatial heterogeneity is also known as structural instability. For linear models, this implies that the linear relationship changes geographically over space, and the linear regression coefficients may form subgroups. It is an important and challenging problem to identify the correct grouping structure of the regression coefficients, as only a correct model structure can lead to unbiased estimation of the regression coefficients and their valid inference. In practice, ad hoc grouping of states as regions defined by tradition or for federal administrative purposes are sometimes used to address this issue. However, such grouping is not driven by the data in specific problems, and may not be appropriate or efficient. For example, the central region in Figure \ref{fig_mapest} (a) includes not only all the Mountain states, but also the North and South Dakotas which are not traditionally considered Mountain states. Figure \ref{fig_mapest} (b) suggests further division of the east to sub-regions, which does not align to any well known administrative regions.    
One natural approach is to assume that regression coefficients of states nearby are more likely to belong to the same subgroup than states which are further apart, and use both the estimated regression coefficients and the spatial structure to guide the clustering of states into subregions, which is what we propose to do in this paper.

The problem of taking into account spatial dependence structure in linear regression has been studied for a long time in literature. Classical works include spatial expansion methods \citep{casetti1972generating,casetti1987spatial}, which treat the spatially varying regression coefficients as a function of expansion variables, typically using longitude and latitude coordinates as location variables. One popular approach to accounting for spatial variations in the model is by introducing  an additive spatial random effect for each location, as done for linear models by \cite{cressie2015statistics} and generalized linear models by \cite{diggle1998model}. Another class of models in wide use are spatial varying coefficient models, including the geographically weighted regression (GWR; \citealt{brunsdon1996geographically}) and its extensions to generalized linear models \citep{nakaya2005geographically} and the Cox model for survival analysis \citep{xue2020geographically,hu2020modified}. There has been also development in the Bayesian framework, such as \cite{gelfand2003spatial}.

The methods mentioned above typically assume that the regression parameters are smooth functions of location variables, which is reasonable in certain practice, but may not be appropriate for applications where 
the covariate effects are constant over subregions defined by some unobserved hidden factors. In this work, we take a different perspective by grouping the covariate effects into spatially-interpretable subgroups or clusters. As in the motivating example, different clusters have different patterns, which can be used to build more flexible estimators to improve the original direct estimates. 
A majority of existing work in the literature on spatial cluster detection are based on hypothesis tests, including the scan statistic methods based on the likelihood ratio \citep{kulldorff1995spatial,jung2007spatial,cook2007spatial} and the two-step spatial test methods under the GWR framework \citep{lee2017cluster,lee2020spatial}. Test-based methods are intuitive and useful in practice, but proper test statistics are often difficulty to construct and the tests may have low power when the number of locations is large. In addition, these methods handle the cluster detection problem and the model estimation separately, making it difficult to study inferential properties of the final estimator. The main purpose of this article is to fill this gap by developing a unified framework to detect clusters of regression coefficients, estimate them consistently, and make valid inferences. 

In the context of non-spatial data analysis, a variety of clustering methods have been proposed to identify homogeneous groups for either observations or regression coefficients.  \cite{chi2015splitting} developed a method for the convex clustering problem through the alternating direction method of multiplier algorithm (ADMM) \citep{boyd2011distributed} with pairwise $L_p (p\geq 1)$ penalty. Nonnegative weights are considered to reduce bias for pairwise penalties. \cite{fan2017l_0} considered a clustering problem with $l_0$ penalty on graphs. For clustering the regression coefficients,  \cite{ma2017concave} and \cite{ma2016estimating} proposed a concave fusion approach for estimating the group structure and estimating subgroup-specific effects, where smoothly clipped absolute deviation (SCAD) penalty \citep{fan2001variable} and the minimax concave penalty (MCP) \citep{zhang2010nearly} are considered.

For spatial analysis, some interesting work are recently proposed for grouping regression coefficients in the spatial regression.  \cite{ma2020heterogeneous} proposed the Bayesian heterogeneity pursuit regression models to detect clusters in the covariate effects based on the Dirichlet process. \cite{hu2020bayesian} proposed a Bayesian method for clustering coefficients with auxiliary covariates random effects, based on a mixture of finite mixtures (MFM).  \cite{li2019spatial} proposed a penalized approach based on the minimum spanning tree.  In the area of spatial boundaries detection, \cite{lu2005bayesian} and \cite{lu2007bayesian} considered the areal boundary detection using a Bayesian hierarchical model based on the conditional autoregressive model \citep{banerjee2014hierarchical}. The boundaries were determined by the posterior distribution of the corresponding spatial process or spatial weights. These boundary detection methods  focused on clustering of observations instead of regression coefficients.

The main difficulty in clustering spacial covariate effects is how to estimate the number of clusters and cluster memberships consistently, by taking into account spatial neighborhood information properly. To our knowledge, none of the existing methods can address this challenge. In this work, we fill the gap by proposing a new procedure, called Spatial Heterogeneity Automatic Detection and Estimation (SHADE), for automatically grouping and estimating local covariate effects simultaneously. The SHADE employs a class of spatially-weighted fusion type penalty on all pairs of observations, with location-specific weight adaptively constructed using geographical proximity of locations, and achieves spatial clustering consistency for spatial linear regression.  In theory,  we show that the oracle estimator based on weighted least squares is a local minimizer of the objective function with probability approaching 1 under certain regular conditions, which indicates that the the number of clusters can be estimated consistently.
 We believe that this result is the first of its kind in the contest of spatial data analysis. To implement the SHADE,  an alternating direction method of multiplier algorithm (ADMM) algorithm is developed. To make the best choices of spatial weights and understand their roles in practical applications, we consider a number of different schemes to choose pairwise weights and compare them numerically and theoretically. Our numerical examples suggest that,  the number of clusters and the group structure can be recovered with high probability,  and the spatial information can help in spatial clustering analysis when the minimal group difference is small or the number of repeated measures is small. 

The article is organized as follows. In Section \ref{sec_model}, we describe the Spatial Heterogeneity Automatic Detection and Estimation (SHADE) model and the corresponding ADMM algorithm.  In Section \ref{sec_thm}, we establish theoretical properties of the SHADE estimator. The simulation study is conducted in Section \ref{sec_sim} under several scenarios to show performances of the proposed estimator. The proposed method is applied to an NRI small area estimation problem to illustrate the use of SHADE in real-data applications in Section \ref{sec_example}. Finally, some discussions are given in Section \ref{sec_dis_cluster}.

\section{Methodology and Algorithm}
\label{sec_model}

\subsection{Methodology: SHADE} 
\label{subsec:model}

Assume our spatial data consist of multiple measurements at each location or subject. Let $y_{ih}$ be the $h$th response for the $i$th subject observed at location $\bm{s}_{i}$,  where $i=1,\dots,n,\,h=1,\dots,n_{i}$. Based on their effects on the response variable, the covariates can be divided into two categories: ``global'' covariates which have common effects on the response across all the locations, and ``local'' covariates which have location-specific effects on the response. To reflect this, let $\bm{z}_{ih}$ and $\bm{x}_{ih}$ be the corresponding covariate vectors with dimension $q$ and $p$ respectively, where $\bm{z}_{ih}$'s are ``global'' covariates which have common linear effects to the response across all the locations, while $\bm{x}_{ih}$'s are ``local'' covariates which have location-specific linear effects on the response. We consider the following linear regression model 
\begin{equation}
\label{eq_model}
y_{ih}=\bm{z}_{ih}^{T}\bm{\eta}+\bm{x}_{ih}^{T}\bm{\beta}_{i}+\epsilon_{ih},
\end{equation}
 where $\bm{\eta}$ represents the vector of common regression coefficients shared by global effects, $\bm{\beta}_i$'s are location-specific regression coefficients, and $\epsilon_{ih}$'s are i.i.d random errors with $E\left(\epsilon_{ih}\right)=0$ and $Var\left(\epsilon_{ih}\right)=\sigma^{2}$. Furthermore, some locations may have same or similar location-specific effects, grouping locations of same location-specific effects can help to achieve dimension reduction and improve model prediction accuracy. Assume the $n$ location-specific effects belong to $K$ mutually exclusive subgroups:  the locations with a common $\bm{\beta}_i$ belong to the same group. Denote the corresponding partition of 
$\{1,\dots,n \}$ as $\mathcal{G}=\left\{ \mathcal{G}_{1},\dots,\mathcal{G}_{K}\right\}$,  where the index set $G_k$ contains all the locations belonging to the group $k$ for $k=1,\ldots, K$.  For convenience, denote the regression coefficients  associated with  $\mathcal{G}_k$ as $\bm{\alpha}_k$.  In practice, since neither $K$ nor the partition $\mathcal{G}_k$'s are known, the goal is to  use the observed data $\{(y_{ih}, \bm{z}_{ih}, \bm{x}_{ih})\}$ to construct the estimator $\hat{K}$ and the partition $\hat{\mathcal{G}} = \{ \hat{\mathcal{G}}_1,\dots, \hat{\mathcal{G}}_{\hat{K}} \}$, where  $\hat{\mathcal{G}}_k =\{i: \hat{\bm{\beta}}_i = \hat{\bm{\alpha}}_k, 1\leq i \leq n \}$. 

To achieve this goal, we use the following optimization problem: minimize the  weighted least squares objective function  subject to a spatially-weighted pairwise penalty
\begin{equation}
\label{eq_obj}
Q_{n}\left(\bm{\eta},\bm{\beta} ; \lambda, \psi\right)=\frac{1}{2}\sum_{i=1}^{n}\frac{1}{n_{i}}\sum_{h=1}^{n_{i}}\left(y_{ih}-\bm{z}_{ih}^{T}\bm{\eta}-\bm{x}_{ih}^{T}\bm{\beta}_{i}\right)^{2}+\sum_{1\leq i<j\leq n}p_{\gamma}\left(\left\Vert \bm{\beta}_{i}-\bm{\beta}_{j}\right\Vert ,c_{ij}\lambda\right),
\end{equation}
where $\bm{\eta}=(\eta_1,\ldots, \eta_q)^T$, $\bm{\beta} = \left( \bm{\beta}_1^T,\dots, \bm{\beta}_n^T\right)^T$, $\Vert \cdot \Vert$ denotes the Euclidean norm, $p_{\gamma}\left(\cdot,\lambda\right)$ is a penalty function imposed on all distinct pairs. In the penalty function, $\lambda \geq 0$ is a tuning parameter, $\gamma>0$ is a built-in constant in the penalty function, and different weights $c_{ij}$'s are assigned to different pairs of locations $\bm{s}_i$ and $\bm{s}_j$  for any $1\le i<j\le n$. One popular choice of penalty is the $L_1$ penalty (lasso) \citep{tibshirani1996regression} with the form $p_{\gamma}(t,\lambda)  = \lambda|t|$. Since $L_1$ penalty tends to produce too many groups as shown in \cite{ma2017concave}, we consider the SCAD penalty, which is defined as
\begin{equation}
\label{eq_scad}
p_{\gamma}(t,\lambda) = \lambda \int_0^{|t|} \min \{ 1,(\gamma - x/\lambda)_+/(\gamma - 1) \}dx.
\end{equation}
Here we treat $\gamma$ as a fixed value as in \cite{fan2001variable}, \cite{zhang2010nearly} and \cite{ma2016estimating}.

\subsection{Choices of Spatial Weights}
\label{subsec:weight}

In \eqref{eq_obj}, the values weights $c_{ij}$ are crucial, as they control the number of subgroups and grouping results. The pairs $\left \Vert \bm{\beta}_{i}-\bm{\beta}_{j}\right\Vert$ with larger weights $c_{ij}\lambda$ are shrunk together more than those pairs with smaller weights. For spatial data, reasonable choices of $c_{ij}$ should take into account two factors: locations with closer $\bm{\beta}_j$ values are more grouped together, and locations closer to each other are more likely to form a subgroup as they typically have similar trends. Since the true values of $\bm{\beta}$ are not available, we use their estimators $\tilde{\bm{\beta}}$ as the surrogates. For example, we can define the weights $c_{ij}$ as
 \[
c_{ij}=\exp\left(-\psi\left\Vert \bm{s}_{i}-\bm{s}_{j}\right\Vert \cdot\left\Vert \tilde{\bm{\beta}}_{i}-\tilde{\bm{\beta}}_{j}\right\Vert \right), 
\]
where $\tilde{\bm{\beta}}_{i}$ is an initial estimate of $\bm{\beta}_i$, and $\psi$ is a  scale parameter  to control magnitudes of the weights. In areal data, we  suggest three different ways of taking into spatial information in the data to construct the weights.\\
(i) using  both spatial and regression coefficients information:  
\begin{equation}
\label{eq_weight}
c_{ij}=\exp\left(\psi\left(1-a_{ij}\right)\cdot\left\Vert \tilde{\bm{\beta}}_{i}-\tilde{\bm{\beta}}_{j}\right\Vert \right),
\end{equation}
where $a_{ij}$ is the neighbor order between location $\bm{s}_i$ and location $\bm{s}_j$, which means that if $i$ and $j$ are neighbors, $a_{ij}  = 1$. If $i$ and $j$ are not neighbors, but they have at least one same neighbor, $a_{ij} = 2$. Similarly, we can have all the neighborhood order for all subjects or locations.\\
 (ii) using regression coefficients information only:
\begin{equation}
\label{eq_reg}
c_{ij}=\exp\left(-\psi\left\Vert \tilde{\bm{\beta}}_i-\tilde{\bm{\beta}}_{j}\right\Vert\right).
\end{equation}
(iii) using spatial information only:
\begin{equation}
\label{eq_spatial}
c_{ij}=\exp\left(\psi (1-a_{ij})\right).
\end{equation}

Weights in \eqref{eq_weight} and \eqref{eq_reg} both includes the regression coefficients, which would depend on the accuracy of $\tilde{\bm{\beta}}_i$. If the number of repeated measures is not large, the values of $\tilde{\bm{\beta}}_i$ will not show the real relationship between different locations, which would lead to very bad weights. The phenomenon can be observed in the simulation study. The weights we use here are three special cases which use the information of either regression coefficients or the spatial neighborhood orders. Definitely, there are other ways to construct weights, such as using distance to borrow the spatial information. At least the weights satisfy condition (C4) in Section \ref{sec_thm}, the theoretical results will hold under other conditions. For example, the weights in \eqref{eq_reg} will satisfy (C4) automatically if $\tilde{\bm{\beta}_i}$'s are consistent estimators. That is, besides the condition (C4), there are no other conditions about the format of the weight function.  

\subsection{Algorithm for SHADE}
\label{subsec:alg}

In this section, we describe the ADMM algorithm to solve \eqref{eq_obj} in Section \ref{subsec:model}. The algorithm shares the same spirit as \pcite{ma2017concave}, where a non-weighted penalty is used in non-spatial settings  without repeated measures.

 There are two tuning parameters, $\lambda$ and $\psi$, in the proposed method. We choose them adaptively using some tuning procedures discussed at the end of this section. For now, we fix them and present the computational algorithm for solving \eqref{eq_obj}. Denote the solution as
\begin{equation}
\label{eq_solution}
\left(  \hat{\bm{\eta}}, \hat{\bm{\beta}} \right) = \argmin_{\bm{\eta} \in \mathbb{R}^q, \bm{\beta} \in \mathbb{R}^{np} } Q_n\left(\bm{\eta}, \bm{\beta}, \lambda, \psi \right).
\end{equation}

 First, we introduce the slack variables for all the pairs ($i,j$) $\bm{\delta}_{ij}=\bm{\beta}_{i}-\bm{\beta}_{j}$,  for $1\le i<j\le n$. Then the problem is equivalent to minimizing the following objective function with regard to $(\bm{\eta},\bm{\beta},\bm{\delta})$,
\begin{align*}
 \min_{\bm{\eta},\bm{\beta},\bm{\delta}} ~~ L_{0}\left(\bm{\eta},\bm{\beta},\bm{\delta}\right) & =\frac{1}{2}\sum_{i=1}^{n}\frac{1}{n_{i}}\sum_{h=1}^{n_{i}}\left(y_{ih}-\bm{z}_{ih}^{T}\bm{\eta}-\bm{x}_{ih}^{T}\bm{\beta}_{i}\right)^{2}+\sum_{1\leq i<j\leq n}p_{\gamma}\left(\left\Vert \bm{\delta}_{ij}\right\Vert ,c_{ij}\lambda\right),\\
& \text{subject to} ~~ \bm{\beta}_{i}-\bm{\beta}_{j}-\bm{\delta}_{ij}=\bm{0}, ~~  1\le i<j\le n,
\end{align*}
where $\bm{\delta}=(\bm{\delta}^T_{ij},1\le i<j\le n)^T$.  To handle the equation constraints in the optimization problem, we introduce the augmented Lagrangian
\[
L\left(\bm{\eta},\bm{\beta},\bm{\delta},\bm{v}\right)=L_{0}\left(\bm{\eta},\bm{\beta},\bm{\delta}\right)+\sum_{i<j}\left\langle \bm{v}_{ij},\bm{\beta}_{i}-\bm{\beta}_{j}-\bm{\delta}_{ij}\right\rangle +\frac{\vartheta}{2}\sum_{i<j}\left\Vert \bm{\beta}_{i}-\bm{\beta}_{j}-\bm{\delta}_{ij}\right\Vert ^{2},
\]
where $\bm{v}=(\bm{v}^T_{ij},1\le i<j\le n)^T$  are Lagrange multipliers and $\vartheta>0$ is the penalty parameter. 

 To solve the problem, we use an iterative algorithm which updates $\bm{\beta}, \bm{\eta}, \bm{\delta}, \bm{v}$ sequentially, one at a time. At the $(m+1)$th iteration, given their current values $(\bm{\beta}^{(m)}, \bm{\eta}^{(m)}, \bm{\delta}^{(m)}, \bm{v}^{(m)})$, the updates of $\bm{\eta},\bm{\beta},\bm{\delta},\bm{v}$ are
\begin{align}
\label{eq_v}
\left(\bm{\eta}^{(m+1)},\bm{\beta}^{(m+1)}\right) & =\argmin_{\bm{\eta},\bm{\beta}}L\left(\bm{\eta},\bm{\beta},\bm{\delta}^{{(m)}},\bm{v}^{{(m)}}\right),\nonumber\\
\bm{\delta}^{(m+1)} & =\argmin_{\bm{\delta}}L\left(\bm{\eta}^{(m+1)},\bm{\beta}^{(m+1)},\bm{\delta},\bm{v}^{(m)}\right), \nonumber\\
\bm{v}_{ij}^{(m+1)} & =\bm{v}_{ij}^{m}+\vartheta\left(\bm{\beta}_{i}^{(m+1)}-\bm{\beta}_{j}^{(m+1)}-\bm{\delta}_{ij}^{(m+1)}\right).
\end{align}

To update $\bm{\eta}$ and $\bm{\beta}$, we minimize the following objective function
\[
f\left(\bm{\beta},\bm{\eta}\right)=\left\Vert \bm{\Omega}^{1/2}\left(\bm{y}-\bm{Z}\bm{\eta}-\bm{X}\bm{\beta}\right)\right\Vert ^{2}+\vartheta\left\Vert \bm{A}\bm{\beta}-\bm{\delta}^{(m)}+\vartheta^{-1}\bm{v}^{(m)}\right\Vert ^{2},
\]
where $\bm{y}=\left(y_{11},\dots,y_{1n_{1}},\dots,y_{n1},\dots,y_{n,n_{n}}\right)^{T}$, $\bm{Z}=\left(\bm{z}_{11},\dots,\bm{z}_{1n_{1}},\dots,\bm{z}_{n1},\dots,\bm{z}_{n,n_{n}}\right)^{T}$, $\bm{X}=\text{diag}\left(\bm{X}_{1},\dots,\bm{X}_{n}\right)$ with $\bm{X}_{i}=\left(\bm{x}_{i1},\dots,\bm{x}_{i,n_{i}}\right)^{T}$, $\bm{\Omega}=\text{diag}\left(1/n_{1}\bm{I}_{n_{1}},\dots,1/n_{n}\bm{I}_{n_{n}}\right)$ and $\bm{A}=\bm{D}\otimes\bm{I}_{p}$ with an $[n\left(n-1\right)/2]\times n$ matrix $\bm{D}=\left\{ \left(\bm{e}_{i}-\bm{e}_{j}\right), i<j\right\} ^{T}$, where $\bm{e}_{i}$ is an $n\times1$ vector with $i$th element 1 and other elements 0. Then the solutions for $\bm{\beta}$ and $\bm{\eta}$ are

\begin{equation}
\label{eq_beta}
\bm{\beta}^{(m+1)}=\left(\bm{X}^{T}\bm{Q}_{Z,\Omega}\bm{X}+\vartheta\bm{A}^{T}\bm{A}\right)^{-1}\left[\bm{X}^{T}\bm{Q}_{Z,\Omega}\bm{y}+\vartheta\text{vec}\left(\left(\bm{\Delta}^{(m)}-\vartheta^{-1}\bm{\Upsilon}^{(m)}\right)\bm{D}\right)\right],
\end{equation}

\begin{equation}
\label{eq_eta}
\bm{\eta}^{(m+1)}=\left(\bm{Z}^{T}\bm{\Omega}\bm{Z}\right)^{-1}\bm{Z}^{T}\bm{\Omega}\left(\bm{y}-\bm{X}\bm{\beta}^{(m+1)}\right),
\end{equation}
where $\bm{\Delta}^{(m)}=\left(\bm{\delta}_{ij}^{(m)},i<j\right)_{p\times n\left(n-1\right)/2}$, $\bm{\varUpsilon}^{(m)}=\left(\bm{v}_{ij}^{(m)},i<j\right)_{p\times n\left(n-1\right)/2}$ and 
\[
\bm{Q}_{Z,\Omega}=\bm{\Omega}-\bm{\Omega}\bm{Z}\left(\bm{Z}^{T}\bm{\Omega}\bm{Z}\right)^{-1}\bm{Z}^{T}\bm{\Omega}.
\]


To update $\bm{\delta}_{ij}$'s componentwisely, it is equivalent to  minimize  the following objective function 
\[
\frac{\vartheta}{2}\left\Vert \bm{\varsigma}_{ij}^{(m)}-\bm{\delta}_{ij}\right\Vert ^{2}+p_{\gamma}\left(\left\Vert \bm{\delta}_{ij}\right\Vert ,c_{ij}\lambda\right),
\]
where $\bm{\varsigma}_{ij}^{(m+1)}=\left(\bm{\beta}_{i}^{(m+1)}-\bm{\beta}_{j}^{(m+1)}\right)+\vartheta^{-1}\bm{v}_{ij}^{(m)}$. The solution  based on SCAD penalty  has a closed-form solution as
\begin{equation}
\label{eq_scad_sol}
\bm{\delta}_{ij}^{(m+1)}=\begin{cases}
S\left(\bm{\varsigma}_{ij}^{(m+1)},\lambda c_{ij}/\vartheta\right) & \text{if }\left\Vert \bm{\varsigma}_{ij}^{(m+1)}\right\Vert \le\lambda c_{ij}+\lambda c_{ij}/\vartheta, \\
\frac{S\left(\bm{\varsigma}_{ij}^{(m+1)},\gamma\lambda c_{ij}/\left(\left(\gamma-1\right)\vartheta\right)\right)}{1-1/\left(\left(\gamma-1\right)\vartheta\right)} & \text{if }\lambda c_{ij}+\lambda c_{ij}/\vartheta<\left\Vert \bm{\varsigma}_{ij}^{(m+1)}\right\Vert \le\gamma\lambda c_{ij},\\
\bm{\varsigma}_{ij}^{(m+1)} & \text{if }\left\Vert \bm{\varsigma}_{ij}^{(m+1)}\right\Vert >\gamma\lambda c_{ij},
\end{cases}
\end{equation}
where $\gamma>c_{ij}+c_{ij}/\vartheta$ and $S\left(\bm{w},t\right)=\left(1-t/\left\Vert \bm{w}\right\Vert \right)_{+}\bm{w}$, and $(t)_+  = t$ if $t >0$, 0 otherwise.


\medskip
 In summary, the computational algorithm can be  described as follows.
\begin{algorithm}[H]
	\caption*{{\bf{Algorithm}:} ADMM algorithm}
	\begin{algorithmic}[1]
	    \REQUIRE: Initialize $\bm{\beta}^{(0)}$, $\bm{\delta}^{(0)}$ and $\bm{v}^{(0)}$. 
	    \FOR {$m=0,1,2,\dots$}
		\STATE Update $\bm{\beta}$ by \eqref{eq_beta}.
		\STATE Update $\bm{\eta}$ by \eqref{eq_eta}.
		\STATE Update $\bm{\delta}$ by \eqref{eq_scad_sol} 
		\STATE Update $\bm{v}$ by \eqref{eq_v}.
		\IF{convergence criterion is met} 
		\STATE{Stop and get the estimates}
		\ELSE
		\STATE { $m=m+1$}
		\ENDIF
		\ENDFOR
	\end{algorithmic}
\end{algorithm}

If the size of $n_i$ is reasonable, such as 10 or larger,  we construct the initial values $\tilde{\bm{\beta}}^{(0)}$  by fitting a linear regression model $y_{ih} = \bm{z}_{ih}^T\bm{\eta}+ \bm{x}_{ih}^T\bm{\beta}_i + \epsilon_{ih}$ for each $i = 1,\dots, n$. Then, set $\bm{\delta}_{ij}^{(0)} = \bm{\beta}_i^{(0)} - \bm{\beta}_j^{(0)}$ and $\bm{v}^{(0)} = \bm{0}$. If $n_i = 1$ or small, the initial values can be set using the procedure in \cite{ma2016estimating}.  They used a ridge fusion criterion with a small value of the tuning parameter, then the initial group structure was obtained by assigning objects into $K^*$(a given value) groups by ranking the estimated $\bm{\beta}_i$.

If $\hat{\bm{\delta}}_{ij} = \bm{0}$, then the location $i$ and $j$  belong to the same group. Thus, we can obtain the corresponding estimated partition $\hat{\mathcal{G}}$ and the estimated number of groups $\hat{K}(\lambda,\psi)$.  For each group, its group-specific  parameter vector is  estimated as $\hat{\bm{\alpha}}_{k} = 1/\vert \hat{\mathcal{G}}_k \vert \sum_{i \in \hat{\mathcal{G}}_k} \hat{\bm{\beta}}_i$  for $k=1, \ldots, \hat{K}$.

\begin{remark}
If there are no global covariates, the model is simplified as  $y_{ih} = \bm{x}_{ih}^T\bm{\beta}_i + \epsilon_{ih}$.  The algorithm will be simplified, that is, $\bm{Q}_{Z,\Omega}$ will become $\bm{\Omega}$. The model we use in the  application is the simplified model. 
\end{remark}

\begin{remark}
The convergence criterion used is the same as \cite{ma2017concave}, which is based on the primal residual $\bm{r}^{(m+1)} = \bm{A}\bm{\beta}^{(m+1)} - \bm{\delta}^{(m+1)}$. The algorithm is stopped if $\Vert \bm{r}^{(m+1)} \Vert < \varepsilon$, where $\varepsilon$ is a small positive value.
\end{remark}

 We need to select two tuning parameters, $\lambda$ and $\psi$, in the SaSa algorithm. In this paper, we use the modified Bayes Information Criterion (BIC) \citep{wang2007tuning} to determine the best tuning parameters adaptively from the data. In particular, we have 
\begin{equation}
\label{eq_bic}
\text{BIC}\left( \lambda, \psi \right) = \log \left[ \frac{1}{n}\sum_{i=1}^n \frac{1}{n_i} \left( y_{ih} - \bm{z}_{ih}^T \hat{\bm{\eta}}(\lambda, \psi) - \bm{x}_{ih}^T\hat{\bm{\beta}}_i(\lambda, \psi) \right)^2 \right] + C_n\frac{\log n}{n}\left( \hat{K}(\lambda, \psi)p + q \right),
\end{equation}
where $C_n$ is a positive number which can depend on $n$. Here we use $C_n = c_0 \log \left( \log \left(np + q \right) \right)$  following \cite{ma2017concave} with $c_0 = 0.2$. To select $\psi$, we select the best value from a set of candidate values, such as 0.1, 0.5, 1, 3. For each  given $\psi$, we use the warm start and continuation strategy as in \cite{ma2016estimating} to select tuning parameter $\lambda$.  A grid of $\lambda$ is predefined within $[\lambda_{\text{min}},\lambda_{\text{max}}]$. For each $\lambda$, the initial values are the estimated values from the previous estimation. Denote the selected tuning parameters as $\hat{\lambda}$ and $\hat{\psi}$. Correspondingly, the estimated group number is  $\hat{K} (\hat\lambda, \hat\psi)$, and the estimated regression coefficients are $\hat{\bm{\beta}}$ and $\hat{\bm{\eta}}$.

\section{Theoretical Properties of SHADE}
\label{sec_thm}
In this section, we study theoretical properties of the proposed  SHADE estimator.  Assume $\mathcal{G}_k$'s are the true partition of location-specific regression coefficients. Let $|\mathcal{G}_k|$ be the number of subjects in group $\mathcal{G}_k$ for $k=1,\dots, K$, $ \left| \mathcal{G}_{\min} \right |$ and $\left| \mathcal{G}_{\max} \right |$ be the minimum and maximum group sizes, respectively. Let $\widetilde{\bm{W}}$ be an $n\times K$ matrix with element $w_{ik}$ and $w_{ik}=1$ if $i\in\mathcal{G}_{k}$, $w_{ik}=0$, otherwise. Denote $\bm{W}=\widetilde{\bm{W}}\otimes\bm{I}_{p}$, which is an $np \times Kp$ matrix and  $\bm{U}=\left(\bm{Z},\bm{X}\bm{W}\right)$.  Define $\mathcal{M}_{\mathcal{G}}=\{\bm{\beta}\in\mathbb{R}^{np}:\,\bm{\beta}_{i}=\bm{\beta}_{j},\text{ for }i,j\in\mathcal{G}_{k},\,1\leq k\leq K\}$.  Using these notations, we can then express $\bm{\beta}$ as $\bm{\beta}=\bm{W}\bm{\alpha}$ if $\bm{\beta}\in\mathcal{M}_{\mathcal{G}}$, where $\bm{\alpha}=\left(\bm{\alpha}_{1}^{T},\dots,\bm{\alpha}_{K}^{T}\right)^{T}$. For any positive numbers, $x_n$ and $y_n$, $x_n \gg y_n$ means that $x_n^{-1} y_n = o(1)$.  Define the scaled penalty function by
\begin{equation}
\label{eq_scp}
\rho_{\gamma}(t) = \lambda^{-1} p_{\gamma}(t,\lambda).
\end{equation}

Below are our assumptions, where (C1)  and (C3) follow those in \cite{ma2016estimating}. 

\begin{enumerate}[label={(C\arabic*)}]
\item The function $\rho_{\gamma}(t)$ is symmetric, non-decreasing, and concave on $[0,\infty)$. It is constant for $t\geq a\lambda$ for some constant $a>0$, and $\rho_{\gamma}(0) = 0$. Also, $\rho^{\prime}(t)$ exists and is continuous except for a finite number values of $t$ and $\rho^{\prime}(0+) = 1$.

\item There exist finite positive constants $M_1, M_2, M_3 >0$ such that $\left|x_{ih,l}\right|\leq M_{1}$, $\left|z_{ih,l}\right|\leq M_{1}$ for $j = 1,\dots, n_i$ and $i = 1,\dots, n$ and $ M_2 \leq \max_{i} n_i /\min_{i} n_i \leq M_3$. Also, assume that $\lambda_{\text{min}}\left(\bm{U}^{T}\bm{\Omega}\bm{U}\right)\geq C_{1}\left|\mathcal{G}_{\text{min}}\right|$,  $\lambda_{\text{max}}\left(\bm{U}^{T}\bm{\Omega}\bm{U}\right)\leq C_{1}^{\prime}n$ for some constants $0 < C_1 <\infty$ and $0 < C_1^{\prime} <\infty$,  where $\lambda_{\text{min}}$ and $\lambda_{\text{max}}$ are the corresponding minimum and maximum eigenvalues respectively. In addition, assume that $\sup_{i,h}\left\Vert \bm{x}_{ih}\right\Vert \leq C_{2}\sqrt{p}$ and $\sup_{i,h}\left\Vert \bm{z}_{ih}\right\Vert \leq C_{3}\sqrt{q}$ for some constants  $0 < C_{2} < \infty$ and $ 0 < C_{3} < \infty$. 

\item The random error vector $\bm{\epsilon}=\left(\epsilon_{11},\dots,\epsilon_{1n_{1}},\epsilon_{21},\dots,\epsilon_{2n_{2}},\dots,\epsilon_{n1},\dots,\epsilon_{nn_{n}}\right)^{T}$
has sub-Gaussian tails such that $P\left(\left|\bm{a}^{T}\bm{\epsilon}\right|>\left\Vert \bm{a}\right\Vert x\right)\leq2\exp\left(-c_{1}x^{2}\right)$ for any vector $\bm{a}\in\mathbb{R}^{m}$ and $x>0$, where $0<c_{1}<\infty$ and $m=\sum_{i=1}^{n}n_{i}$.

\item The pairwise weights $c_{ij}$'s are bounded away from zero if $i$ and $j$ are in the same group.
\end{enumerate}

 Conditions (C1) and (C3) are common used in high-dimensional penalized regression problems, which are also used in \cite{ma2016estimating}. Condition (C2) is also similar to the condition mentioned in \cite{ma2016estimating}, also includes the bounded conditions for covariates, which are used in \cite{huang2004polynomial}. In general, if the weights functions are not only defined on a finite support, $c_{ij}$'s will satisfy condition (C4).

First, we establish the properties of the oracle estimator, which is  defined as the weighted least squares estimator assuming that the underlying group structure is known.  Specifically, the oracle estimator of $(\bm{\eta}, \bm{\alpha})$ is
\begin{align}
\left(\hat{\bm{\eta}}^{or},\hat{\bm{\alpha}}^{or}\right) & =\argmin_{\bm{\eta}\in\mathbb{R}^{q},\bm{\alpha}\in\mathbb{R}^{Kp}}\frac{1}{2}\left\Vert \bm{\Omega}^{1/2}\left(\bm{y}-\bm{Z}\bm{\eta}-\bm{X}\bm{W\alpha}\right)\right\Vert ^{2} \nonumber \\
 & =\left(\bm{U}^{T}\bm{\Omega}\bm{U}\right)^{-1}\bm{U}^{T}\bm{\Omega}\bm{y}. \label{eq_oracle}
\end{align}
 And the corresponding oracle estimator of $\bm{\beta}$ is $\hat{\bm{\beta}}^{or} = \bm{W} \hat{\bm{\alpha}}^{or}$. Let $\bm{\alpha}_k^0$ be the true coefficient vector for group $k$, $k=1,\dots, K$ and $\bm{\alpha}^0 = ((\bm{\alpha}_1^0)^T,\dots, (\bm{\alpha}_K^0)^T)^T$, and let $\bm{\eta}^0$ be the true common coefficient vector. The following theorem shows the properties of the oracle estimator.

\begin{theorem}
\label{subgroup_them1}
Suppose 
\[
\left|\mathcal{G}_{\min}\right|\gg\left(q+Kp\right)^{1/2} \max \left(\sqrt{\frac{n}{\min_{i}n_{i}}\log n},\left(q+Kp\right)^{1/2}\right).
\]
Under conditions (C1)-(C3), $q = o(n)$ and $Kp = o(n)$, we have with probability at least $1-2(q+Kp)n^{-1}$,

\[
\left\Vert \left(\begin{array}{c}
\hat{\bm{\eta}}^{or}-\bm{\eta}^{0}\\
\hat{\bm{\alpha}}^{or}-\bm{\alpha}^{0}
\end{array}\right)\right\Vert \leq \phi_n,
\]
and 
\[
\left\Vert \hat{\bm{\beta}}^{or}-\bm{\beta}^{0}\right\Vert  \leq \sqrt{\left|\mathcal{G}_{\max}\right|} \phi_n ;\, \sup_{i}\left\Vert \hat{\bm{\beta}}_{i}^{or}-\bm{\beta}_{i}^{0}\right\Vert  \leq \phi_n, 
\]
where 
\[
\phi_{n}=c_{1}^{-1/2}C_{1}^{-1}M_{1}\sqrt{q+Kp}\left|\mathcal{G}_{\min}\right|^{-1}\sqrt{\frac{n}{\min n_{i}}\log n}.
\]

Furthermore, for any vector $\bm{a}_n \in \mathbb{R}^{q+Kp}$, we have as $n\rightarrow \infty$
\begin{equation}
\label{eq_clt_est}
\sigma_n(\bm{a}_n)^{-1} \bm{a}_{n}^{T}\left(\left(\hat{\bm{\eta}}^{or}-\bm{\eta}^{0}\right)^{T},\left(\hat{\bm{\alpha}}^{or}-\bm{\alpha}^{0}\right)^{T}\right)^{T} \overset{d} \rightarrow N(0,1),
\end{equation}
where 
\begin{equation}
\sigma_n(\bm{a}_n) = \sigma \left[ \bm{a}_{n}^{T}\left(\bm{U}^{T}\bm{\Omega}\bm{U}\right)^{-1}\bm{U}^{T}\bm{\Omega}\bm{\Omega}\bm{U}\left(\bm{U}^{T}\bm{\Omega}\bm{U}\right)^{-1}\bm{a}_{n} \right]^{1/2}.
\end{equation}
\end{theorem}

\begin{remark}
We don't have any specific assumptions about $n_i$. If $\min n_{i}\ll\frac{n}{q+Kp}\log n$, or $\min n_{i}=O\left(\frac{n}{q+Kp}\log n\right)$,
 we have $\left|\mathcal{G}_{\min}\right|\gg\left(q+Kp\right)^{1/2}\sqrt{\frac{n}{\min n_{i}}\log n}$.
If $\min n_{i}\gg\frac{n}{q+Kp}\log n$, we have $\left|\mathcal{G}_{\min}\right|\gg q+Kp$. In this case, if $q$, $p$ and $K$ are fixed values,  what we need is only $1/\left|\mathcal{G}_{\min} \right | = o(1)$.
\end{remark}

\begin{remark}
The model considered in \cite{ma2016estimating} is a special case of our model, and their condition is a special case of our condition, that is when $n_i = 1$.
\end{remark}

\begin{remark}
 If let $\left|\mathcal{G}_{\min} \right| = \delta n/K$ for some constant $0 < \delta \leq 1$, then
\[
\phi_n = c_1^{-1/2}C_1^{-1} M_1 \delta^{-1}K\sqrt{q+Kp}\sqrt{\log n /(n \min n_i)}.
\] Moreover, if $q$, $p$ and $K$ are fixed values, then $\phi_n = C^*\sqrt{\log n /(n \min n_i)}$ for some constant $0<C^* <\infty$.
\end{remark}

Next, we study the properties of our proposed estimator. Let 
\begin{equation}
\label{eq_mindiff}
b_{n}=\min_{i\in\mathcal{G}_{k},j\in\mathcal{G}_{k^{\prime}}}\left\Vert \bm{\beta}_{i}^{0}-\bm{\beta}_{j}^{0}\right\Vert =\min_{k\neq k^{\prime}}\left\Vert \bm{\alpha}_{k}^{0}-\bm{\alpha}_{k^{\prime}}^{0}\right\Vert 
\end{equation}
be the minimal difference among different groups.

\begin{theorem}
\label{subgroup_them2}
Suppose the conditions of Theorem \ref{subgroup_them1} hold and $(C4)$ holds. If $b_n >a\lambda$ and $\lambda \gg \phi_n$ for some constant $a >0$, then there exists a local minimizer $\left(\hat{\bm{\eta}}(\lambda, \psi)^T, \hat{\bm{\beta}}(\lambda, \psi) ^T \right)^T$ of the objective function $Q_n(\bm{\eta},\bm{\beta})$ given in \eqref{eq_obj} such that
\begin{equation}
P\left(\left(\hat{\bm{\eta}}(\lambda, \psi)^T, \hat{\bm{\beta}}(\lambda, \psi) ^T \right)^T = \left((\hat{\bm{\eta}}^{or})^T, (\hat{\bm{\beta}}^{or})^T \right)^T \right) \rightarrow 1.
\end{equation}
\end{theorem}

\begin{remark}
Theorem \ref{subgroup_them2} implies that true group structure can be recovered with probability approaching 1. It also implies that the estimated number of groups $\hat{K}$ satisfies
$P\left( \hat{K} (\lambda, \psi) = K \right) \rightarrow 1.$
\end{remark}

Let $\hat{\bm{\alpha}}(\lambda,\psi )$ be the distinct group vectors of $\hat{\bm{\beta}}(\lambda, \psi)$. According to Theorem \ref{subgroup_them1} and Theorem \ref{subgroup_them2}, we have the following result.
\begin{corollary}
Suppose the conditions in Theorem \ref{subgroup_them2} hold,  for any vector $\bm{a}_n \in \mathbb{R}^{q+Kp}$, we have as $n\rightarrow \infty$
\begin{equation}
\label{eq_clt_alp}
\sigma_n(\bm{a}_n)^{-1} \bm{a}_{n}^{T}\left(\left(\hat{\bm{\eta}}(\lambda,\psi)-\bm{\eta}^{0}\right)^{T},\left(\hat{\bm{\alpha}}(\lambda,\psi)-\bm{\alpha}^{0}\right)^{T}\right)^{T} \overset{d} \rightarrow N(0,1).
\end{equation}
\end{corollary}

\begin{remark}
The variance parameter $\sigma^2$ can be estimated by
\begin{equation}
\label{eq_sig2}
\sigma^2 = \frac{1}{m -  q - \hat{K}p}\sum_{i=1}^n \sum_{h=1}^{n_i} \left( y_{ih} - \bm{z}_{ih}^T\hat{\bm{\eta}} - \bm{x}_{ih}^T \hat{\bm{\beta}}_i \right)^2
\end{equation}
\end{remark}

The algorithm can be implemented through package \emph{Spgr} in \url{https://github.com/wangx23/Spgr}.

\section{Simulation Studies}
\label{sec_sim}

In this section, we evaluate and compare performance of the proposed  SHADE estimator with different weight choices: equal weights $c_{ij} = 1$ (denoted as  ``equal''), weights defined in \eqref{eq_weight} (denoted as  ``reg-sp''), weights defined in \eqref{eq_reg} (denoted by  ``reg''), and weights defined in \eqref{eq_spatial} (denoted by  ``sp'').

The simulations are carried as follows. Let $\bm{z}_{ih}=(z_{ih,1},\dots,z_{ih,5})^{T}$ with $z_{ih,1}=1$ and $(z_{ih,2},\dots,z_{ih,5})^{T}$ are generated a multivariate normal distribution with mean 0, variance 1, and pairwise correlation $\rho=0.3$. Define $\bm{x}_{ih}=(x_{ih,1},x_{ih,2})^T$, where $x_{ih,1}$ is simulated from a standard normal distribution and $x_{ih,2}$ is simulated from a centered and standardized binomial $(n,0.7)$. Let $\bm{\eta}=(\eta_{1},\dots,\eta_{5})^{T}$, where  $\eta_{k}$'s are simulated from Uniform $[1,2]$ and standard deviation of the error term is $\sigma=0.5$. We set $\vartheta=1$ and $\gamma=3$ and use the SCAD penalty function. The tuning parameters are chosen by the modified BIC defined by \eqref{eq_bic}. We consider the simulations in  several scenarios. The results are based on 100 simulations.

To evaluate subgrouping performance of the proposed method, we report the estimated
group number $\hat{K}$, adjusted Rand index (ARI) \citep{rand1971objective, hubert1985comparing, vinh2010information}, and the root mean square error (RMSE) for estimating $\bm{\beta}$. For the estimated $\hat{K}$ over 100 simulations, we report its average (denoted by  ``mean''),  standard error in the parenthesis, and the occurrence percentage of $\hat{K} = K$ (denoted by  ``per''). The quantity ARI is used to measure the degree of agreement between two partitions, taking a value between 0 and 1: the larger ARI value, the more agreement.  We report the average ARI across 100 simulations along with the standard error in the  parentheses. To evaluate estimation accuracy of $\bm{\beta}$,  we also report the average RMSE
\begin{equation}
\label{eq_RMSE}
\sqrt{\frac{1}{n}\sum_{i=1}^n\Vert \hat{\bm{\beta}}_i - \bm{\beta}_i \Vert^2}.
\end{equation}

\subsection{Balanced group}
We assume that there are  $K=3$ true groups $\mathcal{G}_{1},\mathcal{G}_{2}$ and $\mathcal{G}_{3}$. Consider the two spatial settings,  for which the group parameters are respectively given by:

Setting 1: $\bm{\beta}_{i}=(1,1)^{T}$ if $i\in\mathcal{G}_{1}$; $\bm{\beta}_{i}=(1.5,1.5)^{T}$ if $i\in\mathcal{G}_{2}$; $\bm{\beta}_{i}=(2,2)^{T}$ if $i\in\mathcal{G}_{3}$. 

Setting 2:  $\bm{\beta}_{i}=(1,1)^{T}$ if $i\in\mathcal{G}_{1}$; $\bm{\beta}_{i}=(1.25,1.25)^{T}$ if $i\in\mathcal{G}_{2}$; $\bm{\beta}_{i}=(1.5,1.5)^{T}$ if $i\in\mathcal{G}_{3}$. 

\begin{figure}[H]
\centering
\begin{subfigure}{.4\textwidth}
  \centering
  \includegraphics[width=0.9\linewidth]{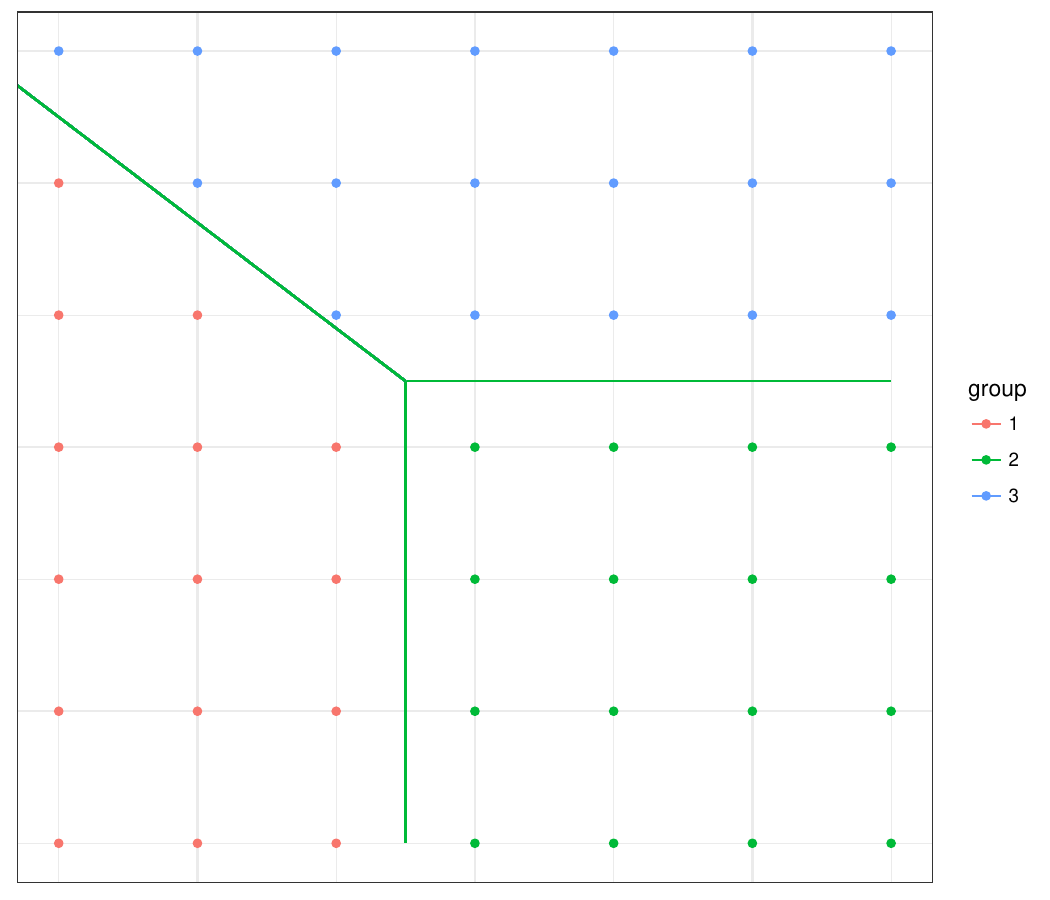}
  \caption{$7\times 7$ grid}
  \label{fig_grid49}
\end{subfigure}%
\begin{subfigure}{.4\textwidth}
  \centering
  \includegraphics[width=0.9\linewidth]{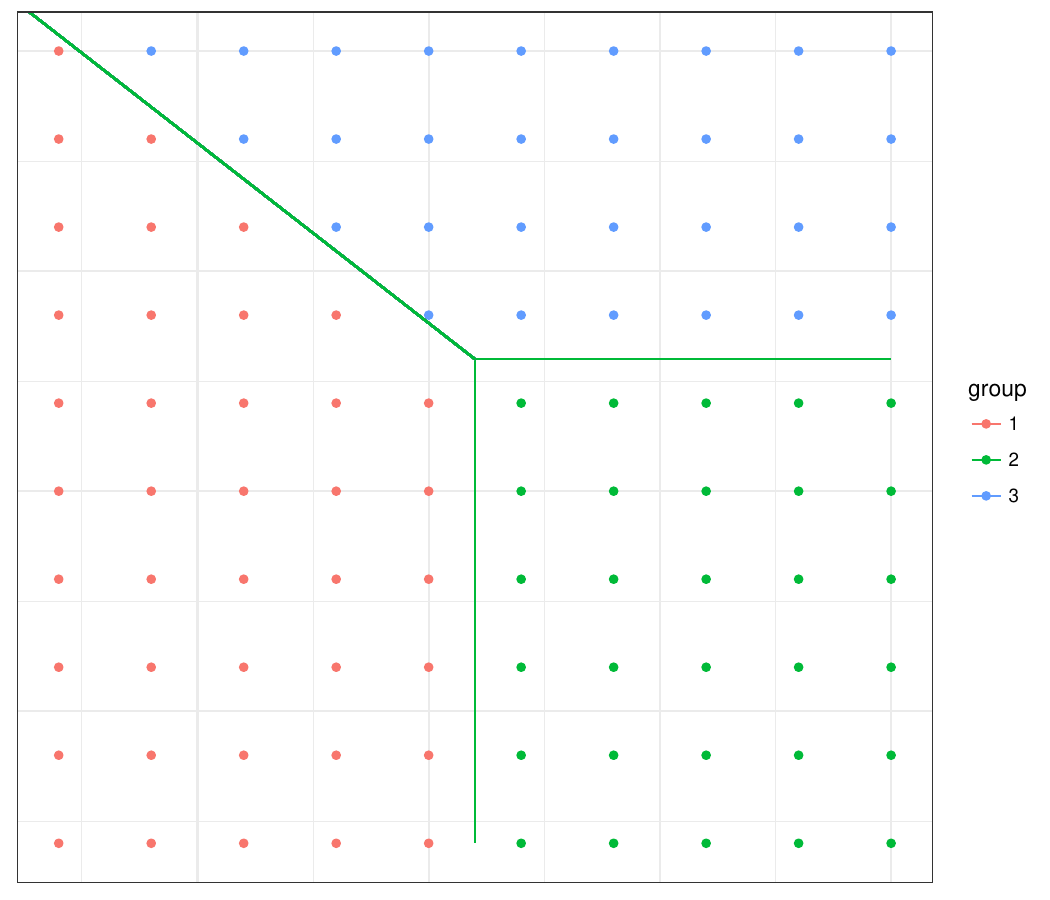}
  \caption{$10\times 10$ grid}
  \label{fig_grid100}
\end{subfigure}
\caption{Two spatial settings in the simulation studies.}
\label{fig_grid}
\end{figure}

\noindent
Under each setting, we simulate the data on two sizes of regular lattice, a $7\times 7$ grid  (left) and a $10\times 10$ grid  (right), as shown in Figure \ref{fig_grid}.  Furthermore, for the $7\times 7$ grid  with $n_i=10$, we  use a 10-fold cross validation to select the tuning parameters.  The repeated measures of location $i$ are divided into 10 parts;  the $j$th part of each location is combined as the validation data set, the remaining observations form the training data set.  The spatial weights \eqref{eq_spatial}  are considered. The results are labeled as ``cv" in all the tables.  Note that ``reg\_sp" and ``reg" were not computed for the $10\times 10$ grid.

\begin{figure}[H]
\begin{center}
\includegraphics[scale=0.6]{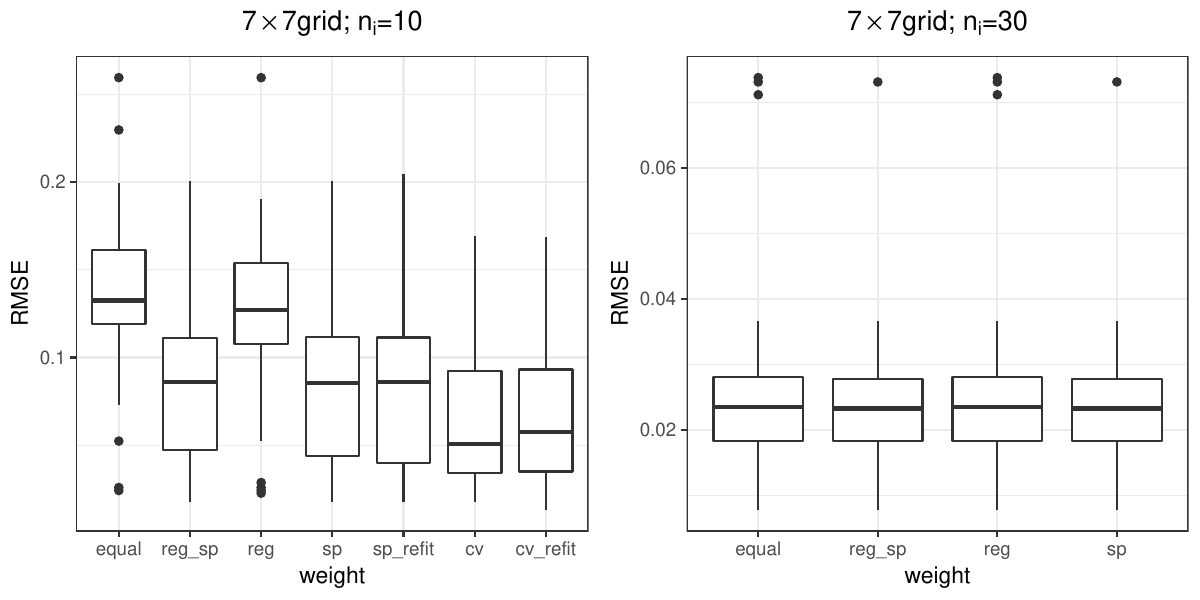}
\caption{RMSE for Setting 1 under the $7\times 7$ grid}
\label{fig_RMSE_grid49_p1}
\end{center}
\end{figure}

\begin{figure}[H]
\begin{center}
\includegraphics[scale=0.6]{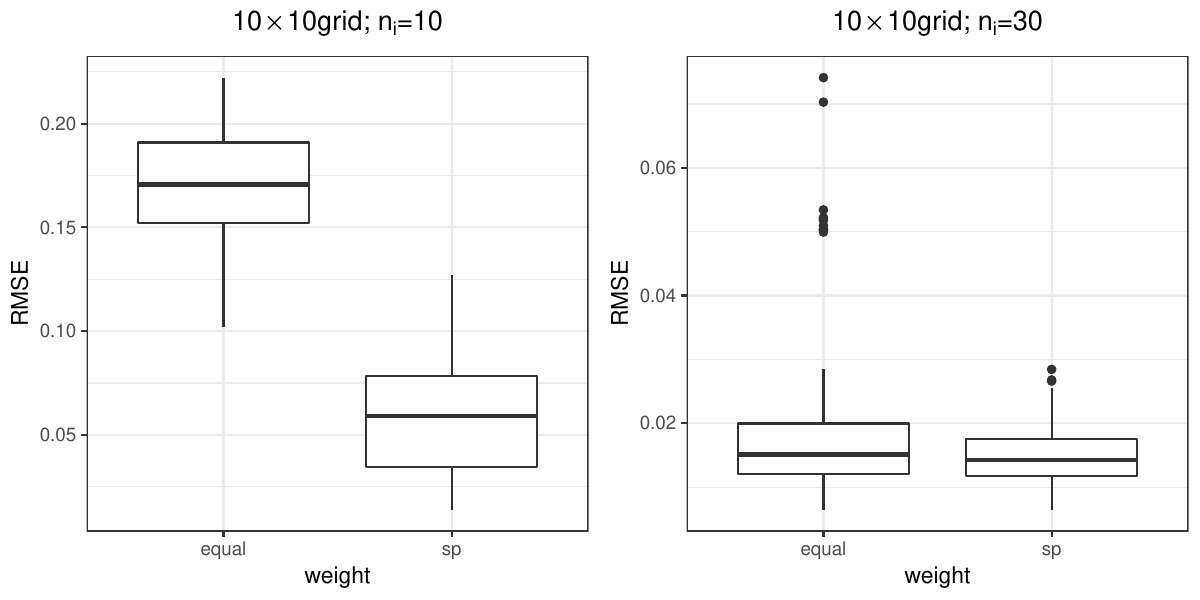}
\caption{RMSE for Setting 1 under the $10\times 10$ grid}
\label{fig_RMSE_grid100_p1}
\end{center}
\end{figure}

\noindent {\bf Results for Setting 1}: 
Tables \ref{tab_K_grid49_p1} , \ref{tab_R_grid49_p1} and  \ref{tab_KR_grid100_p1} show the  estimated number of groups and ARI. Figures \ref{fig_RMSE_grid49_p1}  and \ref{fig_RMSE_grid100_p1}  plot the RMSE of the estimates obtained using different weight choices. After estimating the group structure, one can also estimate parameters $\bm{\eta}$ and $\bm{\beta}$ again by assuming that the group information is known; the results are denoted as  ``refit''. Based on the numerical results, we make the following observations.

First, we summarize the results for the $7\times 7$ grid. In all the  considered scenarios,  the spatially weighted penalty outperforms the non-weighed penalty ( ``equal''). The upper panels in Tables
\ref{tab_K_grid49_p1} and \ref{tab_R_grid49_p1}, the left plot in Figure  \ref{fig_RMSE_grid49_p1} suggest that, if the number of repeated measurements is relatively small (say, $n_i = 10$), the weights  ``reg\_sp'' and ``sp'' perform similarly and they are the best in terms of estimating $K$, recovering the true subgroup structure (large ARI), and estimating regression coefficients (small RMSE); the weights  ``equal'' and ``reg'' are
much worse. The lower panels of Tables \ref{tab_K_grid49_p1}  and \ref{tab_R_grid49_p1}  and the right plot in Figure  \ref{fig_RMSE_grid49_p1} show that, when the number of repeated measurements gets larger (say, $n_i = 30$), all the methods improve and there is not much difference among them. Cross validation works well in terms of ARI and RMSE, but it tends to over-estimate the number of groups $K$. This is because that cross validation focuses more on the prediction accuracy; the coefficient estimates of some groups are close to the true  coefficients, but they are not shrank together. In addition, refitting the model does not appear to further improve the accuracy of estimating $\bm{\beta}$.

\begin{table}[H]
\centering
\caption{Summary of the estimate $\hat{K}$ for Setting 1 under the $7\times7$ grid.}
\label{tab_K_grid49_p1}
\begin{tabular}{cc|ccccc}
  \hline
 && equal & reg\_sp & reg & sp & cv  \\ 
  \hline
 \multirow{ 2}{*}{$n_i=10$} &  mean & 3.34(0.054) & 3.15(0.039) & 3.33(0.051) & 3.13(0.034) & 3.82(0.13)\\
  & per & 0.69 & 0.86 & 0.69 & 0.87 & 0.56 \\
   \midrule
   \multirow{ 2}{*}{$n_i=30$} & mean & 3.00(0) & 3.00(0) & 3.00(0) & 3.00(0) \\
   & per & 1.00 & 1.00 & 1.00 & 1.00 \\ 
   \hline
\end{tabular}
\end{table}

\begin{table}[H]
\centering
\caption{Average ARI for Setting 1 under the $7\times7$  grid}
\label{tab_R_grid49_p1}
\begin{tabular}{c|ccccc}
  \hline
 & equal & reg\_sp & reg & sp & cv \\ 
  \hline
  $n_i=10$ & 0.80(0.011) & 0.92(0.008) & 0.82(0.01) & 0.92(0.007) & 0.95(0.007) \\
  \midrule
  $n_i=30$ & 0.998(0.001) & 0.999(0.0006) & 0.998(0.001) & 0.999(0.0006) \\ 
   \hline
\end{tabular}
\end{table}

Next, we summarize the results for the $10\times10$ grid.   In this case, we consider equal weights and spatial weights only. Again,  the spatially-weighted penalty outperforms the non-weighed penalty ( ``equal''). Table \ref{tab_KR_grid100_p1} and Figure \ref{fig_RMSE_grid100_p1} suggest that, if the number of repeated measurements is relatively small (say, $n_i = 10$),  ``sp'' performs much better in terms of grouping and estimating regression coefficients than ``equal''; for a larger number of repeated measurements (say, $n_i =  30$), they perform similarly.

\begin{table}[H]
\centering
\caption{ Summary of $\hat{K}$ and average ARI for Setting 1 under the $10\times10$  grid.}
\label{tab_KR_grid100_p1}
\begin{tabular}{cc|cc|cc}
  \hline
  && \multicolumn{2}{c|}{$\hat{K}$} &  \multicolumn{2}{c}{ARI}\\
   \hline
 && equal & sp & equal & sp  \\ 
  \hline
\multirow{2}{*}{$n_i=10$} & mean & 3.59(0.073) & 3.37(0.065)  & 0.70(0.009) & 0.97(0.003)   \\ 
  &per & 0.53 & 0.71  & -& - \\ 
  \midrule
\multirow{2}{*}{$n_i=30$} & mean &3(0) & 3(0) & 0.996(0.001) & 1.00(0) \\
&per & 1.00 & 1.00 &  - & -\\
   \hline
\end{tabular}
\end{table}

\begin{figure}[H]
\begin{center}
\includegraphics[scale=0.6]{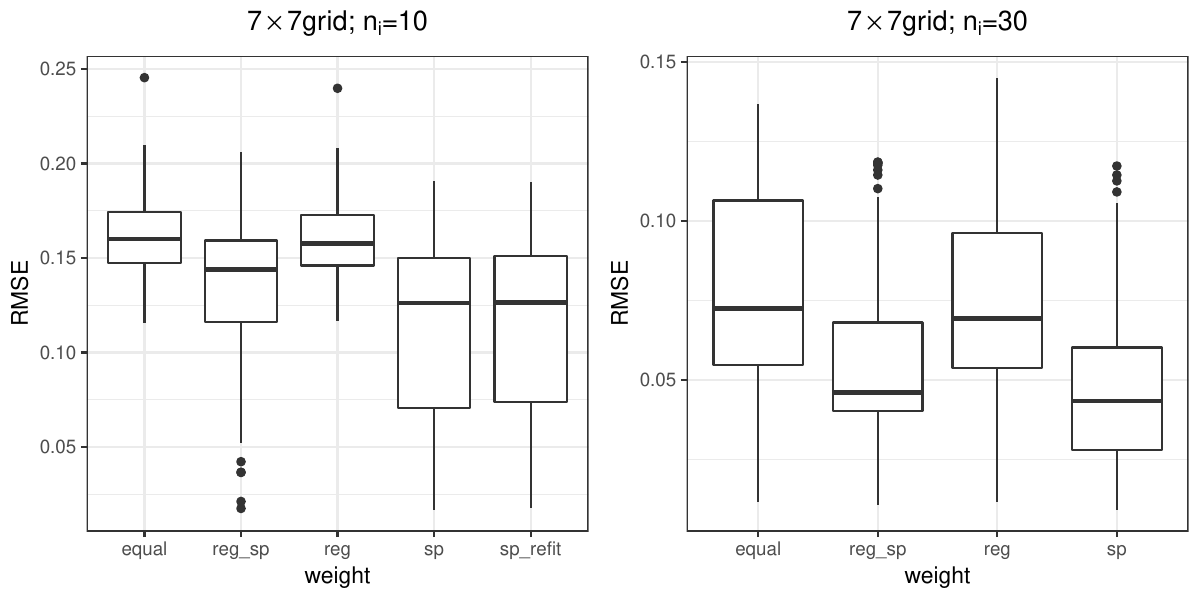}
\caption{RMSE for setting 2 under the $7\times 7$ grid}
\label{fig_RMSE_grid49_p2}
\end{center}
\end{figure}

\noindent {\bf Results for Setting 2}: 
In this setting, the group difference becomes smaller. Tables \ref{tab_K_grid49_p2}, \ref{tab_R_grid49_p2} and Figure \ref{fig_RMSE_grid49_p2} summarize the results for the $7\times 7$ grid.  For both values of $n_i$, the weights ``sp'' performs best in terms of estimating the number of groups ($\hat{K}$), recovering the true group structure (ARI), and estimating regression coefficients.  In contrast to Setting 1, when the difference among groups becomes smaller, even with $n_i = 30$,  the model with the spatial weight is superior to other models.

\begin{table}[H]
\centering
\caption{Summary of $\hat{K}$ for Setting 2 under the $7\times7$  grid}
\label{tab_K_grid49_p2}
\begin{tabular}{cc|cccc}
  \hline
& & equal & reg\_sp & reg & sp   \\ 
  \hline
\multirow{2}{*}{$n_i=10$} & mean & 3.25(0.119) & 3.01(0.093) & 3.14(0.107) & 2.88(0.067) \\
  & per & 0.34 & 0.45 & 0.33 & 0.60 \\
  \midrule
  \multirow{2}{*}{$n_i=30$} & mean & 2.70(0.046) & 2.90(0.030) & 2.76(0.043) & 2.95(0.022) \\ 
  & per & 0.70 & 0.90 & 0.76 & 0.95 \\ 
   \hline
\end{tabular}
\end{table}

\begin{table}[H]
\centering
\caption{ Average ARI for Setting 2 under the $7\times7$  grid}
\label{tab_R_grid49_p2}
\begin{tabular}{c|cccc}
  \hline
 & equal & reg\_sp & reg & sp  \\
  \hline
$n_i = 10$& 0.32(0.011) & 0.50(0.023) & 0.33(0.01) & 0.61(0.026)\\
\midrule 
$n_i = 30$ & 0.72(0.018) & 0.86(0.015) & 0.75(0.017) & 0.90(0.012) \\ 
   \hline
\end{tabular}
\end{table}

\begin{figure}[H]
\begin{center}
\includegraphics[scale=0.6]{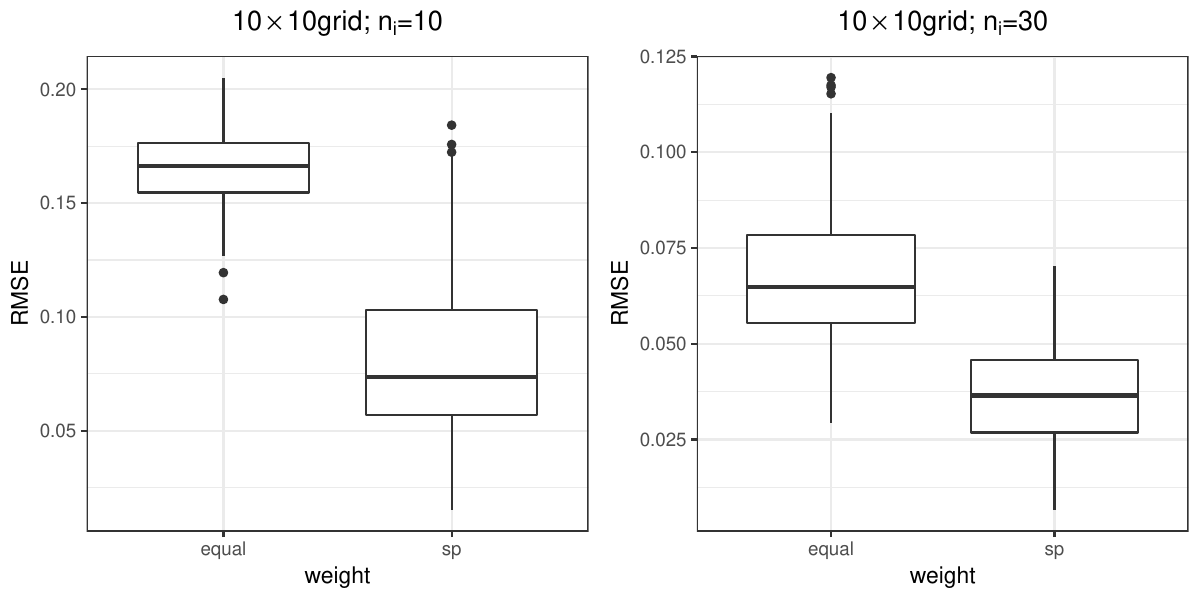}
\caption{RMSE for Setting 2 under the $10\times 10$ grid}
\label{fig_RMSE_grid100_p2}
\end{center}
\end{figure}

Table \ref{tab_KR_grid100_p2} and Figure \ref{fig_RMSE_grid100_p2} show the results for the $10\times 10$ grid. Again, we only compare ``equal'' weights and ``sp'' weights. The results suggest similar conclusions to  those for the $7\times 7$ grid: the model with the spatial weight is superior even with a large number of repeated measurements ($n_i = 30$)  by producing larger ARI and smaller RMSE.

\begin{table}[H]
\centering
\caption{Summary of $\hat{K}$ and average ARI for Setting 2 under the $10\times10$  grid}
\label{tab_KR_grid100_p2}
\begin{tabular}{cc|cc|cc}
  \hline
  & & \multicolumn{2}{c|}{$\hat{K}$} &  \multicolumn{2}{c}{ARI}\\
  \hline 
 & & equal & sp & equal & sp \\
  \hline
 \multirow{2}{*}{$n_i=10$} & mean & 3.82(0.146) & 3.35(0.078)  & 0.32(0.009) & 0.81(0.022) \\
 &  per & 0.32 & 0.620& -& - \\ 
 \midrule
  \multirow{2}{*}{$n_i=30$}& mean &3.10(0.060) & 3.00(0.0) & 0.79(0.012) & 0.94(0.005) \\ 
 & per & 0.64 & 1.0 &-& -\\
   \hline
\end{tabular}
\end{table}

\subsection{Unbalanced group setting}
\begin{figure}[H]
\begin{center}
\includegraphics[scale=0.4]{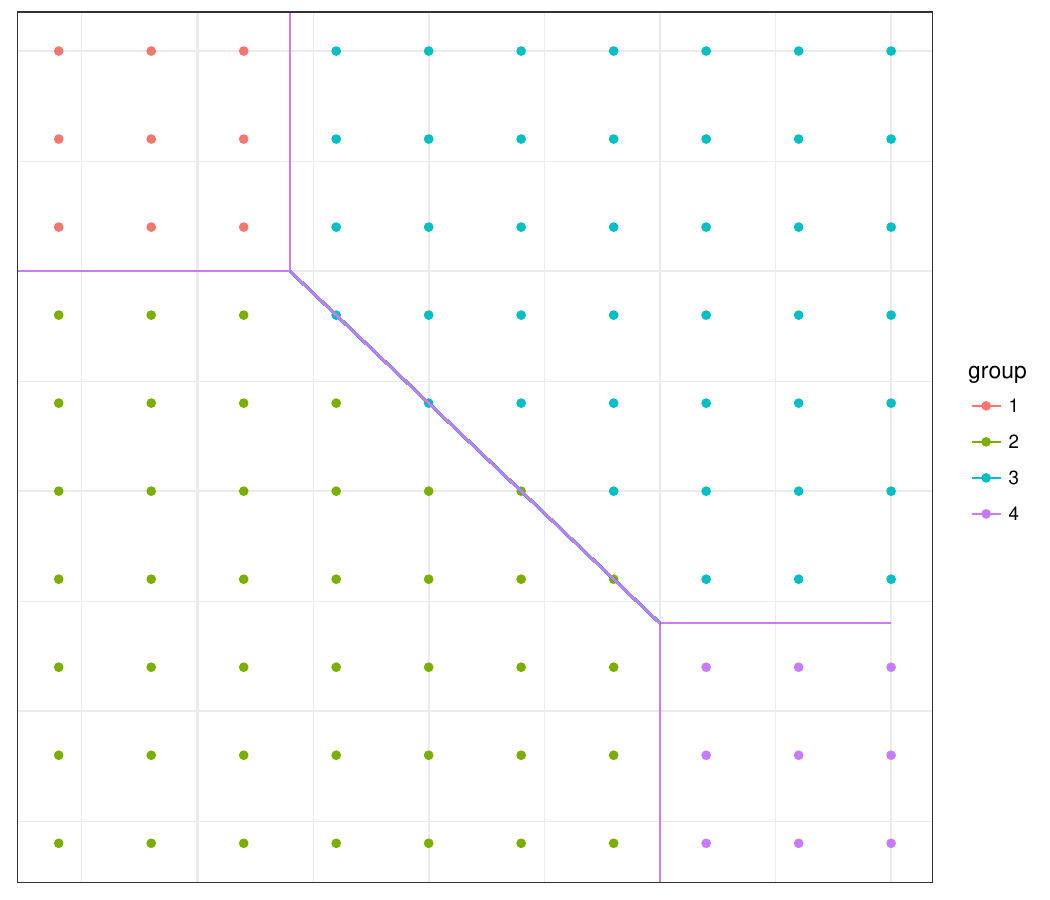}
\caption{Unbalanced group setting}
\label{fig_grid100_ub}
\end{center}
\end{figure}

Here we consider an unbalanced group setting as shown in Figure \ref{fig_grid100_ub}. In this setting, there are four groups, denoted as $\mathcal{G}_1, \mathcal{G}_2, \mathcal{G}_3$ and  $\mathcal{G}_4$,  and two groups have 9 subjects and the other two groups have 41 subjects. The group parameters are $\bm{\beta}_{i}=(1,1)^{T}$ if $i\in\mathcal{G}_{1}$, $\bm{\beta}_{i}=(1.5,1.5)^{T}$ if $i\in\mathcal{G}_{2}$, $\bm{\beta}_{i}=(2,2)^{T}$ if $i\in\mathcal{G}_{3}$ and $\bm{\beta}_{i}=(2.5,2.5)^{T}$ if $i\in\mathcal{G}_{4}$.

\begin{figure}[H]
\begin{center}
\includegraphics[scale=0.6]{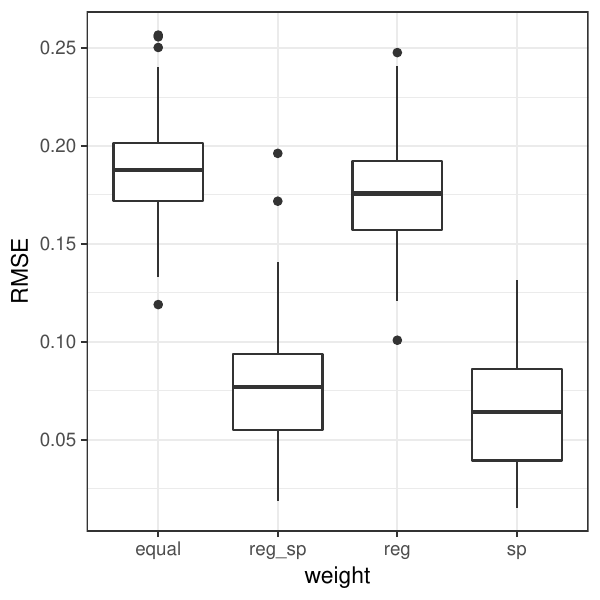}
\caption{RMSE for unbalanced setting}
\label{fig_RMSE_ub}
\end{center}
\end{figure}

Table \ref{tab_ub_p1} and Figure \ref{fig_RMSE_ub} show the summaries of $\hat{K}$, ARI and RMSE for $\bm{\beta}$  when the number of repeated measurements is $n_i = 10$.  Overall speaking,  ``reg\_sp" and ``sp" perform better than the other two types of weights.  Especially, ``sp" performs a slightly better than ``reg\_sp" . The results are consistent with those under balanced cases.  We expect that, when the group difference becomes smaller, ``sp" would still perform better than other weights even when the number of repeated measurements is large. 

\begin{table}[H]
\centering
\caption{Summary of $\hat{K}$ and average ARI for the unbalanced setting with $n_i=10$}\label{tab_ub_p1}
\begin{tabular}{cc|cccc}
  \hline
 && equal & reg\_sp & reg & sp  \\ 
  \hline
\multirow{2}{*}{$\hat{K}$} & mean & 4.58(0.093) & 4.23(0.049) & 5.17(0.011) & 4.35(0.059) \\
  & per & 0.570 & 0.800 & 0.300 & 0.710 \\
  \midrule
 ARI & mean & 0.62(0.010) & 0.94(0.061) & 0.67(0.009) & 0.96(0.004) \\ 
   \hline
\end{tabular}
\end{table}

\subsection{Random group setting}  
\begin{figure}[H]
\begin{center}
\includegraphics[scale=0.6]{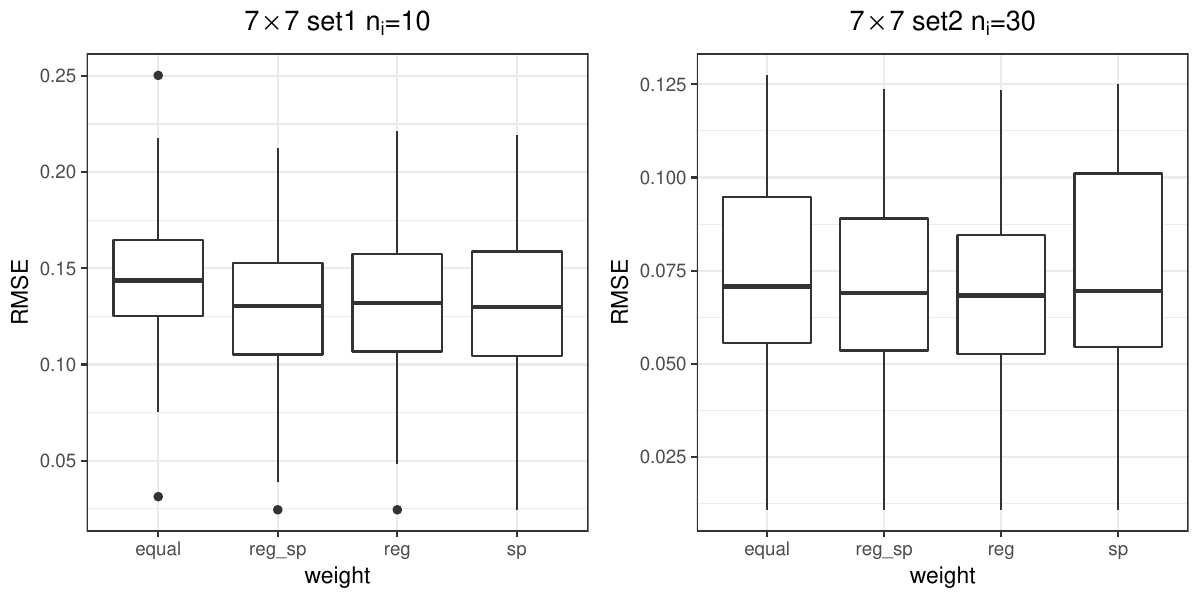}
\caption{RMSE for random groups under the $7\times 7$ grid}
\label{fig_RMSE_grid49_random}
\end{center}
\end{figure}

We consider a setting without specified location group information. For each location, it has equal probability to three groups.  Tables \ref{tab_random_p1} shows the summary of $\hat{K}$ and ARI for Setting 1 under the grid $7\times7$ with 10 repeated measures.  Table  \ref{tab_random_p2}  shows the summary of $\hat{K}$ and ARI for Setting 2 under the grid $7\times7$ with 30 repeated measures.  Figure \ref{fig_RMSE_grid49_random} shows the RMSE results for both cases. We can see that different weights have similar performances. The results suggest that even without prior information on the existence of spatial groups, ``sp'' weights can still produce comparable results as equal weights.

\begin{table}[ht]
\centering
\caption{Summary of $\hat{K}$ and average ARI  for Setting 1 under the $7\times 7$ grid with $n_i=10$}
\label{tab_random_p1}
\begin{tabular}{cc|cccc}
  \hline
 & & equal & reg\_sp & reg & sp\\ 
  \hline
\multirow{2}{*}{$\hat{K}$} & mean & 3.42(0.064) & 3.45(0.063) & 3.40(0.059) & 3.45(0.063)  \\ 
  &per & 0.66 & 0.62 & 0.65 & 0.62 \\ 
  \midrule 
ARI & mean& 0.78(0.011) & 0.82(0.010) & 0.81(0.010) & 0.82(0.011)\\
   \hline
\end{tabular}
\end{table}

\begin{table}[ht]
\centering
\caption{Summary of $\hat{K}$ and average ARI for Setting 2 under the $7\times 7$ grid  with $n_i = 30$}
\label{tab_random_p2}
\begin{tabular}{cc|cccc}
  \hline
 && equal & reg\_sp & reg & sp  \\ 
  \hline
\multirow{2}{*}{$\hat{K}$} & mean & 2.77(0.045) & 2.77(0.045) & 2.83(0.040) & 2.73(0.047) \\
  & per & 0.75 & 0.75 & 0.81 & 0.71  \\ 
  \midrule
  ARI & mean & 0.74(0.015) & 0.76(0.016) & 0.77(0.014) & 0.74(0.017) \\ 
   \hline
\end{tabular}
\end{table}

\section{Application}
\label{sec_example}

In this section, we apply our  SHADE method to the modeling of the National  Resources Inventory survey (NRI) data \footnote{https://www.nrcs.usda.gov/wps/portal/nrcs/main/national/technical/nra/nri/} for the purpose of small area estimation. 
The NRI survey is one of the largest longitudinal natural resource surveys in the U.S., and its national and state level estimates of the status and change of land cover and use and soil erosion have been used by numerous federal, state, and local agencies in the past few decades. In recent years, there is an increasing demand for NRI to provide various county level estimates. These include estimates of different land covers, such as cropland, pasture land, urban and forest. Due to the limitation of sample size, the uncertainty of the NRI direct county level estimates are usually too large for local stakeholders for making policy decisions. To make the county level estimates more useful, it is necessarily to include some auxiliary information and appropriate model to reduce the uncertainty of the estimates. One such set of auxiliary covariates is Cropland Data Layer (CDL), which is based on classification of satelliate image pixels into several mutually exclusive and exhaustive land cover categories. In this section, we model the relationship between the NRI forest proportion and the CDL forest proportion among 48 states. In NRI, forests belonging to federal land, such as national parks, are not included in the forest category. For states with more forest federal land, NRI estimates can be substantially smaller than CDL estimates. Therefore, different states could have different relationship between these two proportions. 

The model we consider is,
\begin{equation}
\label{eq_forest}
y_{ih} = \beta_{0,i} + \beta_{1,i}x_{ih} + \epsilon_{ih}
\end{equation}
where $y_{ih}$ is the NRI forest proportion of the $h$th county in the $i$th state, $x_{ih}$ is the corresponding CDL forest proportion of the $h$th county in the $i$th state, and $\beta_{0,i}$ and $\beta_{1,i}$ are the unknown coefficients. Both $x$ and $y$ are standardized. Instead of using the estimated linear regression coefficients as initial values directly,  we use five sets of initial values  which are simulated from a multivariate normal distribution with estimated coefficients as the mean vector and estimated covariance matrix as the covariance matrix. The models with   the smallest modified BIC values are selected for equal weights and spatial weights respectively.  

\begin{figure}[H]
\centering
\begin{subfigure}[b]{.4\textwidth}
  \centering
  \includegraphics[width=1\linewidth]{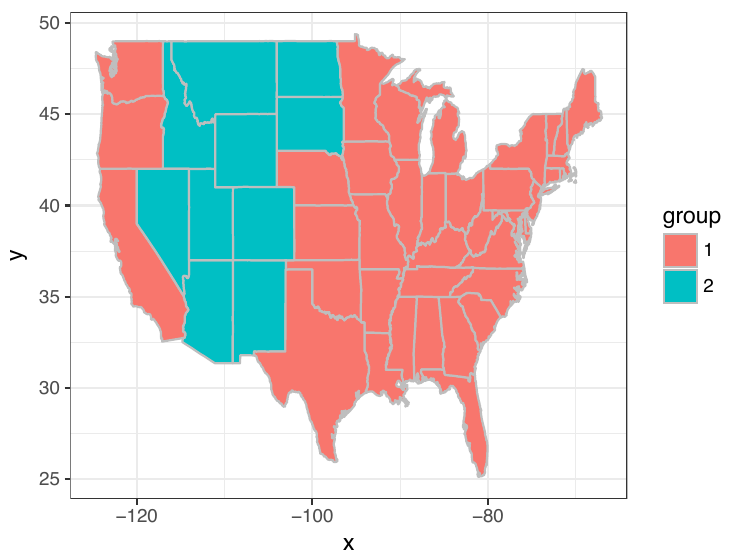}
  \caption{Estimated groups based on equal weights}
  \label{map_equal}
\end{subfigure}%
~
\begin{subfigure}[b]{.4\textwidth}
  \centering
  \includegraphics[width=1\linewidth]{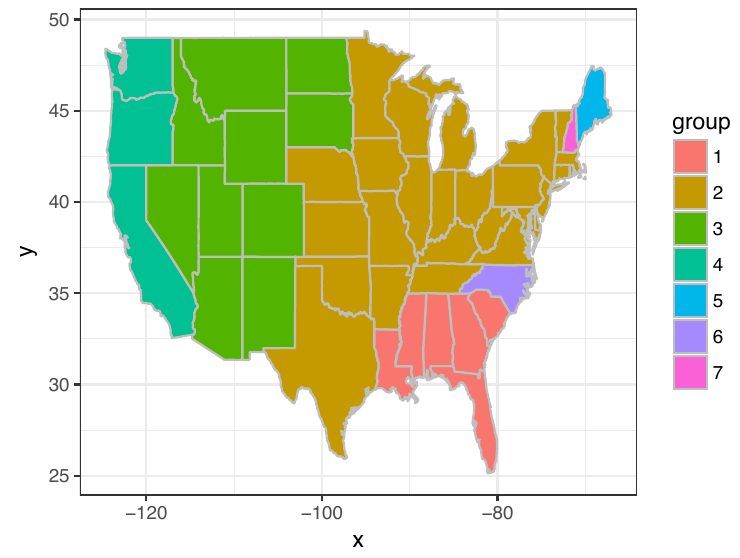}
  \caption{Estimated groups based on spatial weights}
  \label{map_spatial}
\end{subfigure}
\vspace{-0.1in}
\caption{Estimated groups for both equal weights and spatial weights.}
\label{fig_mapest}
\end{figure}

Figure \ref{fig_mapest} shows the estimated groups based on 2011 NRI data sets. The left figure  plots the estimated groups based on equal weights, and the right one is for the estimated groups based on spatial weights in \eqref{eq_spatial}. We find that the two different weights give different estimated groups. Tables \ref{tab_forest_eq} and \ref{tab_forest_sp} are the corresponding estimates of regression coefficients in different groups.  

\begin{table}[H]
\centering
\caption{Estimated coefficients of different groups for equal weight}
\label{tab_forest_eq}
\begin{tabular}{c|cc}
  \hline
group & 1 & 2 \\ 
  \hline
$\beta_0$ & -0.029(0.006) & 0.003(0.008) \\ 
 $\beta _1 $ & 0.885(0.011) & 0.241(0.026) \\ 
   \hline
\end{tabular}
\end{table}

\begin{table}[H]
\centering
\caption{Estimated coefficients of different groups for spatial weights}
\label{tab_forest_sp}
{\scriptsize
\begin{tabular}{c|ccccccc}
  \hline
 group & 1 & 2 & 3 & 4 & 5 & 6 & 7 \\ 
 \hline
$\beta_0$ & -0.041(0.016) & -0.032(0.006) & 0.003(0.007) & 0.023(0.015) & -0.108(0.293) & 0.275(0.038) & 0.376 (0.309) \\ 
 $\beta_1$ & 1.018(0.028) & 0.867(0.012) & 0.241(0.024) & 0.608(0.033) & 1.148 (0.377) & 0.332(0.064) & 0.341(0.384) \\ 
   \hline
\end{tabular}
}
\end{table}

\begin{figure}[H]
\begin{center}
\includegraphics[scale=0.6]{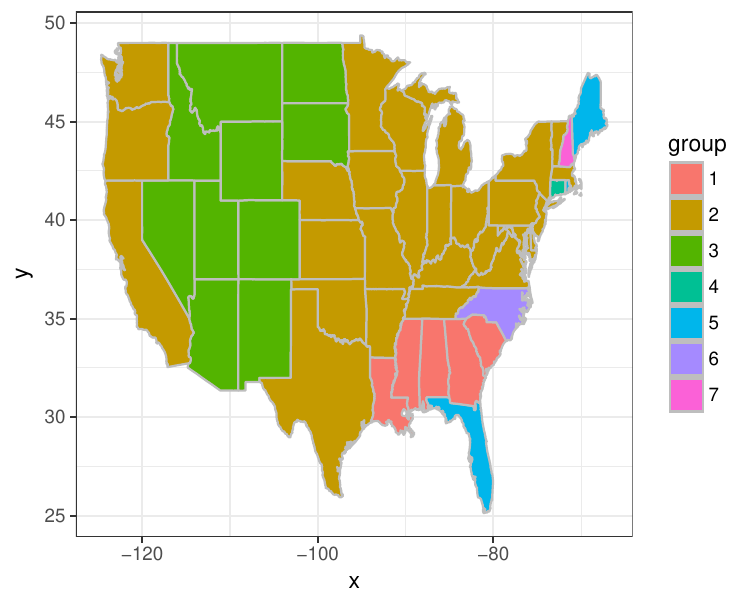}
\caption{Estimated groups by changing the tuning parameter $\lambda$ with equal weights.}
\label{fig_mapest_eq}
\end{center}
\end{figure}

When considering equal weights, $\lambda$ is the only tuning parameter in the algorithm. By changing the value of $\lambda$, we can have different number of groups. We consider to change the $\lambda$ value in the algorithm based on equal weights such that the number of groups is the same as what we have selected based on the spatial weights, that is 7 groups.   Figure \ref{fig_mapest_eq} shows the group structure with 7 groups based on equal weights. In both Figure  \ref{fig_mapest_eq}  and  the left figure of Figure \ref{fig_mapest}, ``WA", ``OR'' and ``CA" are not separated from the majority group (the group with the largest group size) when considering equal weights. These three states are in group 4, which are separated from the majority group (group 2) when considering spatial weights, which is more reasonable and intuitive based on the estimates of  regression coefficients as shown in Table \ref{tab_forest_sp}. Besides that, these three states have more national parks than those states in group 2.

\begin{figure}[H]
\begin{center}
\includegraphics[scale=0.5]{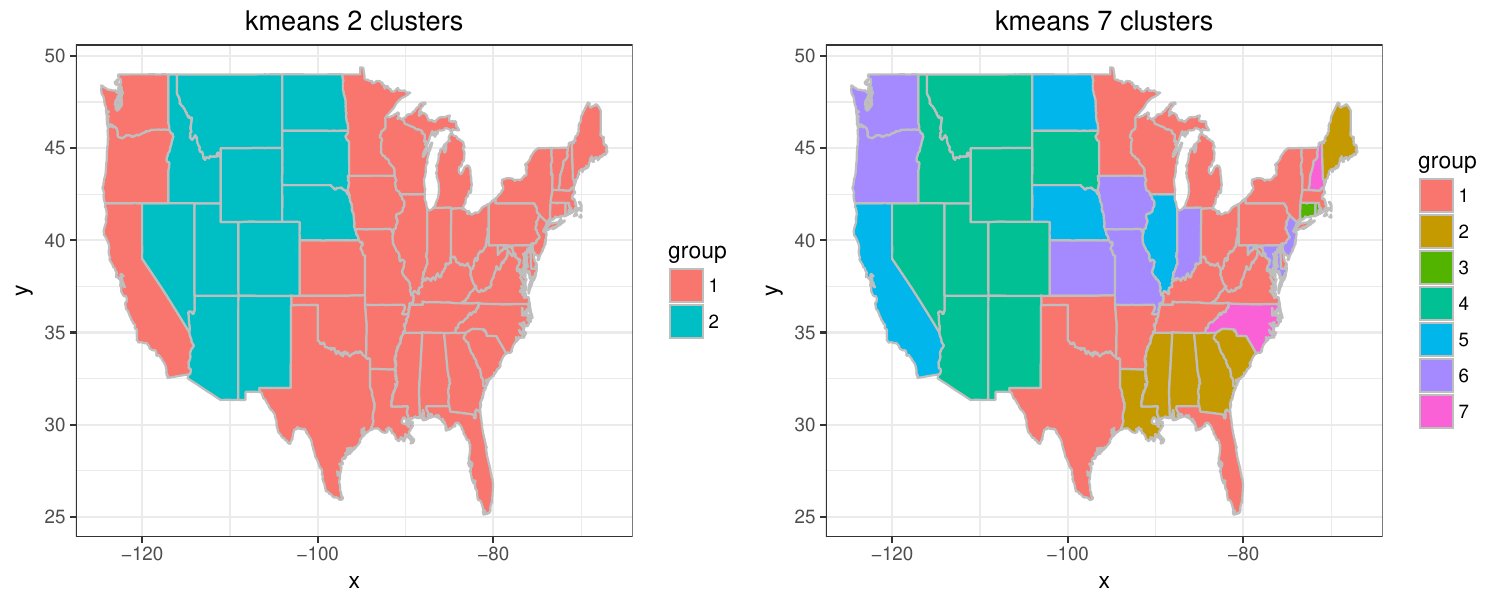}
\caption{Group clustering results based on K-means.}
\label{fig_kmeans}
\end{center}
\end{figure}

Alternatively, we also implement $K$-means clustering based on the initial estimates  to identify similar behaviors among the states. Figure \ref{fig_kmeans} shows the maps based on 2-means clustering and 7-means clustering, respectively. We notice that the 2-cluster map is almost the same as the map based on equal weights. However, the  7-cluster map is not interpretable compared to the result based on spatial weights. This suggests that the proposed procedure can produce more interpretable subgroup structures than $K$-means clustering methods.

\section{Discussion}
\label{sec_dis_cluster}
In this article, we consider the  problem of spatial clustering of local covariate effects and develop a general framework  called Spatial Heterogeneity Automatic Detection and Estimation (SHADE) for spatial areal data with repeated measures. In spatial data, since locations near each other usually have similar patterns,  we propose to take into account spatial information in  the pairwise penalty, where closer locations are assigned with larger weights to encourage stronger shrinkage.  In the simulation study, we use several examples to investigate and compare performance  of the procedure using different weights and have found that spatial information helps to improve the accuracy of grouping, especially when the minimal group difference is small or the number of repeated measures is small. We also establish theoretical properties of the proposed estimator in terms of its consistency in estimating the number of groups. 

In the real data example, we have treated states as locations and counties as repeated measures. Alternatively, one can treat counties as individual units, since one state could have counties with two different features. Then, the algorithm will involve a matrix inverse with dimension more than 3000, which will require higher computational burden. A further study is needed to compare these two models for the application.

The proposed method does not consider the spatial dependence in the regression error when constructing the objective function. The basic idea of this algorithm can be extended to a general spatial clustering setup with consideration of the spatially dependent error. More specifically, the weighted least squared term in the objective function needs to be replaced by a generalized least squares term, which includes an estimated covariance matrix. The new algorithm should have two iterative steps. The first step is to update regression coefficients to find clusters and the second step is to update covariance parameters. More simulation studies are needed to explore the performance of the two-step algorithm. And the theoretical properties needs to be established to support the new algorithm. Both theoretical and computational aspect of such extension are nontrivial and will be considered in a follow up work. 

\begin{appendices}
\renewcommand{\thesubsection}{\Alph{subsection}}
\section*{Appendices}

\subsection{Proof of Theorem \ref{subgroup_them1}}
In this section, we prove Theorem \ref{subgroup_them1}. When proving the central limit theorem (CLT) we use the technique in \cite{huang2004polynomial}.

The oracle estimator is define in \eqref{eq_oracle}, which has the following form
\[
\left(\begin{array}{c}
\hat{\bm{\eta}}^{or}\\
\hat{\bm{\alpha}}^{or}
\end{array}\right)=\left(\bm{U}^{T}\bm{\Omega}\bm{U}\right)^{-1}\bm{U}^{T}\bm{\Omega}\bm{y}.
\]
Thus, we have
\[
\left(\begin{array}{c}
\hat{\bm{\eta}}^{or}-\bm{\eta}^{0}\\
\hat{\bm{\alpha}}^{or}-\bm{\alpha}^{0}
\end{array}\right)=\left(\bm{U}^{T}\bm{\Omega}\bm{U}\right)^{-1}\bm{U}^{T}\bm{\Omega}\bm{\epsilon},
\]
where $\bm{\epsilon} = (\bm{\epsilon}_i^T,\dots, \bm{\epsilon}_n^T)^T$ with $\bm{\epsilon}_i = (\epsilon_{i1},\dots, \epsilon_{i,n_i})^T$. Therefore,
\begin{equation}
\label{eq_normeq}
\left\Vert \left(\begin{array}{c}
\hat{\bm{\eta}}^{or}-\bm{\eta}^{0}\\
\hat{\bm{\alpha}}^{or}-\bm{\alpha}^{0}
\end{array}\right)\right\Vert \leq\left\Vert \left(\bm{U}^{T}\bm{\Omega}\bm{U}\right)^{-1}\right\Vert _2\left\Vert \bm{U}^{T}\bm{\Omega}\bm{\epsilon}\right\Vert, 
\end{equation}
where $\Vert \cdot \Vert_2$ is matrix norm, which is defined as, for a matrix $\bm{A}$, $\Vert \bm{A} \Vert_2 = \sup_{\Vert \bm{x} \Vert  = 1} \Vert \bm{Ax} \Vert $. 

We know that 
\begin{align}
P\left(\left\Vert \bm{U}^{T}\bm{\Omega}\bm{\epsilon}\right\Vert _{\infty}>C\sqrt{\frac{n}{\min n_{i}}\log n}\right) & \leq P\left(\left\Vert \left(\bm{X}\bm{W}\right)^{T}\bm{\Omega}\bm{\epsilon}\right\Vert _{\infty}>C\sqrt{\frac{n}{\min n_{i}}\log n}\right)\nonumber\\
 & +P\left(\left\Vert \bm{Z}^{T}\bm{\Omega}\bm{\epsilon}\right\Vert_{\infty} >C\sqrt{\frac{n}{\min n_{i}}\log n}\right), \label{eq_uinfty}
\end{align}
where $C$ is a finite positive constant and  $\Vert \cdot \Vert_{\infty}$ is defined as, for a vector $\bm{x} \in \mathbb{R}^m$,  $\Vert \bm{x}  \Vert_{\infty} = \max_{1\leq i \leq m} x_i$.  By condition (C2), we have
\[
\sqrt{\sum_{i=1}^{n}\sum_{h=1}^{n_{i}}\frac{x_{ih,l}^{2}}{n_{i}^{2}}1\left\{ i\in\mathcal{G}_{k}\right\} }\leq M_{1}\sqrt{\sum_{i=1}^{n}\frac{1}{n_{i}}\left\{ i\in\mathcal{G}_{k}\right\} }\leq M_{1}\sqrt{\sum_{i=1}^{n}\frac{1}{n_{i}}}\leq M_{1}\sqrt{\frac{n}{\min n_{i}}}.
\]
Since
\[
\left\Vert \left(\bm{X}\bm{W}\right)^{T}\bm{\Omega}\bm{\epsilon}\right\Vert _{\infty}=\sup_{k,l}\left|\sum_{i=1}^{n}\frac{1}{n_{i}}\sum_{h=1}^{n_{i}}x_{ih,l}\epsilon_{ih}1\left\{ i\in\mathcal{G}_{k}\right\} \right|,
\]
from condition (C3), it follows that
\begin{align*}
 & P\left(\left\Vert \left(\bm{X}\bm{W}\right)^{T}\bm{\Omega}\bm{\epsilon}\right\Vert_{\infty} >C\sqrt{\frac{n}{\min n_{i}}\log n}\right)\\
\leq & \sum_{l=1}^{p}\sum_{k=1}^{K}P\left(\left|\sum_{i=1}^{n}\sum_{j=1}^{n_{i}}\frac{1}{n_{i}}x_{ih,l}\epsilon_{ih}1\left\{ i\in\mathcal{G}_{k}\right\} \right|>C\sqrt{\frac{n}{\min n_{i}}\log n}\right)\\
= & \sum_{l=1}^{p}\sum_{k=1}^{K}P\left(\left|\sum_{i=1}^{n}\sum_{h=1}^{n_{i}}\frac{1}{n_{i}}x_{ih,l}\epsilon_{ih}1\left\{ i\in\mathcal{G}_{k}\right\} \right|>\frac{\sqrt{\sum_{i=1}^{n}\sum_{h=1}^{n_{i}}\frac{x_{ih,l}^{2}}{n_{i}^{2}}1\left\{ i\in\mathcal{G}_{k}\right\} }}{\sqrt{\sum_{i=1}^{n}\sum_{h=1}^{n_{i}}\frac{x_{ih,l}^{2}}{n_{i}^{2}}1\left\{ i\in\mathcal{G}_{k}\right\} }}C\sqrt{\frac{n}{\min n_{i}}\log n}\right)\\
\leq & \sum_{l=1}^{p}\sum_{k=1}^{K}P\left(\left|\sum_{i=1}^{n}\sum_{h=1}^{n_{i}}\frac{1}{n_{i}}x_{ih,l}\epsilon_{ih}1\left\{ i\in\mathcal{G}_{k}\right\} \right|>\sqrt{\sum_{i=1}^{n}\sum_{h=1}^{n_{i}}\frac{x_{ih,l}^{2}}{n_{i}^{2}}1\left\{ i\in\mathcal{G}_{k}\right\} }\frac{C}{M_{1}}\sqrt{\log n}\right)\\
\leq & 2Kp\exp\left(-c_{1}\frac{C^{2}}{M_{1}^{2}}\log n\right)=2Kpn^{-c_{1}C^{2}/M_{1}^{2}}.
\end{align*}

Similarly, $\left|\sum_{i=1}^{n}\sum_{h=1}^{n_{i}}\frac{z_{ih,l}^{2}}{n_{i}^{2}}\right|\leq M_{1}^{2}\sum_{i=1}^{n}1/n_{i}\leq M_{1}^{2}\frac{n}{\min n_{i}}$. Again, by condition (C3), we have
\begin{align*}
 & P\left(\left\Vert \bm{Z}^{T}\bm{\Omega}\bm{\epsilon}\right\Vert _{\infty}>C\sqrt{\frac{n}{\min n_{i}}\log n}\right)\\
\leq & \sum_{l=1}^{q}P\left(\left|\sum_{i=1}^{n}\sum_{h=1}^{n_{i}}\frac{1}{n_{i}}z_{ih,l}\epsilon_{ih}\right|>C\sqrt{\frac{n}{\min n_{i}}\log n}\right)\\
\leq & \sum_{l=1}^{q}P\left(\left|\sum_{i=1}^{n}\sum_{h=1}^{n_{i}}\frac{1}{n_{i}}z_{ih,l}\epsilon_{ih}\right|>\sqrt{\sum_{i=1}^{n}\sum_{h=1}^{n_{i}}\frac{z_{ih,l}^{2}}{n_{i}^{2}}}\frac{C}{M_{1}}\sqrt{\log n}\right)\\
\leq & 2q\exp\left(-c_{1}\frac{C^{2}}{M_{1}^{2}}\log n\right)=2qn^{-c_{1}C^{2}/M_{1}^{2}}.
\end{align*}
Thus, \eqref{eq_uinfty} can be bounded by
\[
P\left(\left\Vert \bm{U}^{T}\bm{\Omega}\bm{\epsilon}\right\Vert _{\infty}>C\sqrt{\frac{n}{\min n_{i}}\log n}\right)\leq2\left(Kp+q\right)n^{-c_{1}C^{2}/M_{1}^{2}}. 
\]
Since $\left\Vert \bm{U}^{T}\text{\ensuremath{\bm{\Omega}\bm{\epsilon}}}\right\Vert \leq\sqrt{q+Kp}\left\Vert \bm{U}^{T}\bm{\Omega}\bm{\epsilon}\right\Vert _{\infty}$, 
\[
P\left(\left\Vert \bm{U}^{T}\bm{\Omega}\bm{\epsilon}\right\Vert >C\sqrt{q+Kp}\sqrt{\frac{n}{\min n_{i}}\log n}\right)\leq2\left(Kp+q\right)n^{-c_{1}C^{2}/M_{1}^{2}}.
\]
Let $C=c_{1}^{-1/2}M_{1}$, thus
\begin{equation}
\label{eq_vectornormineq}
P\left(\left\Vert \bm{U}^{T}\bm{\Omega}\bm{\epsilon}\right\Vert >C\sqrt{q+Kp}\sqrt{\frac{n}{\min n_{i}}\log n}\right)\leq2\left(Kp+q\right)n^{-1}.
\end{equation}
Also, according to condition (C2), we have 
\begin{equation}
\label{eq_inverseineq}
\left\Vert \left(\bm{U}^{T}\bm{\Omega}\bm{U}\right)^{-1}\right\Vert_2 \leq C_{1}^{-1}\left|\mathcal{G}_{\text{min}}\right|^{-1}. 
\end{equation}

Combining \eqref{eq_normeq}, \eqref{eq_vectornormineq} and \eqref{eq_inverseineq}, with probability at least $1-2\left(Kp+q\right)n^{-1}$, we have
\[
\left\Vert \left(\begin{array}{c}
\hat{\bm{\eta}}^{or}-\bm{\eta}^{0}\\
\hat{\bm{\alpha}}^{or}-\bm{\alpha}^{0}
\end{array}\right)\right\Vert \leq CC_{1}^{-1}\sqrt{q+Kp}\left|\mathcal{G}_{\text{min}}\right|^{-1}\sqrt{\frac{n}{\min n_{i}}\log n}. 
\]
Let \[\phi_n = c_1^{-1/2}C_{1}^{-1}M_1\sqrt{q+Kp}\left|\mathcal{G}_{\text{min}}\right|^{-1}\sqrt{\frac{n}{\min n_{i}}\log n}.\] Furthermore, 
\begin{align*}
\left\Vert \hat{\bm{\beta}}^{or}-\bm{\beta}^{0}\right\Vert ^{2} & =\sum_{k=1}^{K}\sum_{i\in\mathcal{G}_{k}}\left\Vert \hat{\bm{\alpha}}_{k}^{or}-\bm{\alpha}_{k}^{0}\right\Vert ^{2}\leq\left|\mathcal{G}_{\text{max}}\right|\sum_{k=1}^{K}\left\Vert \hat{\bm{\alpha}}_{k}^{or}-\bm{\alpha}_{k}^{0}\right\Vert ^{2}\\
 & =\left|\mathcal{G}_{\text{max}}\right|\left\Vert \hat{\bm{\alpha}}^{or}-\bm{\alpha}^{0}\right\Vert ^{2}\leq\left|\mathcal{G}_{\text{max}}\right|\phi_{n}^{2},
\end{align*}
and
\[
\sup_{i}\left\Vert \hat{\bm{\beta}}_{i}^{or}-\bm{\beta}_{i}^{0}\right\Vert =\sup_{k}\left\Vert \hat{\bm{\alpha}}_{k}^{or}-\bm{\alpha}_{k}^{0}\right\Vert \leq\left\Vert \hat{\bm{\alpha}}^{or}-\bm{\alpha}^{0}\right\Vert \leq\phi_{n}.
\]

Next, we consider the central limit theorem. Let $\bm{U} = \left( \bm{U}_1^T, \dots, \bm{U}_n^T\right)^T $ with $\bm{U}_i = (\bm{U}_{i1},\dots, \bm{U}_{i,n_i})^T$ for $i=1,\dots,n$. Consider
\begin{align*}
\bm{a}_{n}^{T}\left(\left(\hat{\bm{\eta}}^{or}-\bm{\eta}^{0}\right)^{T},\left(\hat{\bm{\alpha}}^{or}-\hat{\bm{\alpha}}^{0}\right)^{T}\right)^{T}  = &\sum_{i=1}^{n}\bm{a}_{n}^{T}\left(\sum_{i=1}^{n}\bm{U}_{i}^{T}\bm{\Omega}_{i}\bm{U}_{i}\right)^{-1}\bm{U}_{i}^{T}\bm{\Omega}_{i}\bm{\epsilon}_{i},
\end{align*}
where $\bm{\Omega}_i = 1/n_i \bm{I}_{n_i}$. By the assumption of $\bm{\epsilon}_i$ in the model \eqref{eq_model}, we have
\[
E\left[\bm{a}_{n}^{T}\left(\left(\hat{\bm{\eta}}^{or}-\bm{\eta}^{0}\right)^{T},\left(\hat{\bm{\alpha}}^{or}-\hat{\bm{\alpha}}^{0}\right)^{T}\right)^{T}\right]=0.
\]
The variance of $\bm{a}_{n}^{T}\left(\left(\hat{\bm{\eta}}^{or}-\bm{\eta}^{0}\right)^{T},\left(\hat{\bm{\alpha}}^{or}-\hat{\bm{\alpha}}^{0}\right)^{T}\right)^{T}$ can be written as
\begin{align*}
 & Var\left\{ \bm{a}_{n}^{T}\left(\left(\hat{\bm{\eta}}^{or}-\bm{\eta}^{0}\right)^{T}, \left(\hat{\bm{\alpha}}^{or}-\hat{\bm{\alpha}}^{0}\right)^{T}\right)^{T}\right\} \\
= & \sigma^{2}\left[\bm{a}_{n}^{T}\left(\bm{U}^{T}\bm{\Omega}\bm{U}\right)^{-1}\bm{U}^{T}\bm{\Omega}\bm{\Omega}\bm{U}\left(\bm{U}^{T}\bm{\Omega}\bm{U}\right)^{-1}\bm{a}_{n}\right]\\
= & \sigma^{2}\left[\bm{a}_{n}^{T}\left(\bm{U}^{T}\bm{\Omega}\bm{U}\right)^{-1}\sum_{i=1}^{n}\bm{U}_{i}^{T}\bm{\Omega}_{i}\bm{\Omega}_{i}\bm{U}_{i}\left(\bm{U}^{T}\bm{\Omega}\bm{U}\right)^{-1}\bm{a}_{n}\right].
\end{align*}
We use the technique of \cite{huang2004polynomial} in the proof of their Theorem 3. That is,\\$\bm{a}_{n}^{T}\left(\left(\hat{\bm{\eta}}^{or}-\bm{\eta}^{0}\right)^{T},\left(\hat{\bm{\alpha}}^{or}-\hat{\bm{\alpha}}^{0}\right)^{T}\right)^{T}$ can be written as $\sum_{i=1}^{n}a_{i}\xi_{i}$ with 
\[
a_{i}^{2}=\bm{a}_{n}^{T}\left(\bm{U}^{T}\bm{\Omega}\bm{U}\right)^{-1}\bm{U}_{i}^{T}\bm{\Omega}_{i}\bm{\Omega}_{i}\bm{U}_{i}\left(\bm{U}^{T}\bm{\Omega}\bm{U}\right)^{-1}\bm{a_{n}},
\]
where $\xi_{i}$'s are independent with mean zero and variance one. If
\[
\frac{\max_{i}a_{i}^{2}}{\sum_{i=1}^{n}a_{i}^{2}}\rightarrow0,
\]
then $\sum_{i=1}^{n}a_{i}\xi_{i}/\sqrt{\sum_{i=1}^{n}a_{i}^{2}}$ is asymptotically $N\left(0,1\right)$.

For any $\bm{\lambda}=\left(\lambda_{1},\dots,\lambda_{q+Kp}\right)^{T}$, we have
\begin{align*}
\bm{\lambda}^{T}\bm{U}_{i}^{T}\bm{\Omega}_{i}\bm{\Omega}_{i}\bm{U}_{i}\bm{\lambda} & =\frac{1}{n_{i}^{2}}\bm{\lambda}^{T}\bm{U}_{i}^{T}\bm{U}_{i}\bm{\lambda}=\frac{1}{n_{i}^{2}}\sum_{h=1}^{n_{i}}\bm{\lambda}^{T}\bm{U}_{ih}\bm{U}_{ih}^{T}\bm{\lambda}\\
 & =\frac{1}{n_{i}^{2}}\sum_{h=1}^{n_{i}}\left(\sum_{l=1}^{q+Kp}U_{ih,l}\lambda_{l}\right)^{2}\\
 & \leq\frac{1}{n_{i}^{2}}\sum_{h=1}^{n_{i}}\left(\sum_{l=1}^{q+Kp}U_{ih,l}^{2}\right)\left(\sum_{l=1}^{q+Kp}\lambda_{l}^{2}\right)\leq\frac{M_{1}^{2}}{n_{i}}\left(q+Kp\right)\left\Vert \bm{\lambda}\right\Vert ^{2}.
\end{align*}

\begin{align*}
\bm{\lambda}^{T}\left(\sum_{i=1}^{n}\bm{U}_{i}^{T}\bm{\Omega}_{i}\bm{\Omega}_{i}\bm{U}_{i}\right)\bm{\lambda} & \geq\frac{1}{\max_{i}n_{i}}\bm{\lambda}^{T}\left(\sum_{i=1}^{n}\bm{U}_{i}^{T}\bm{\Omega}_{i}\bm{U}_{i}\right)\bm{\lambda}\geq\frac{1}{\max_{i}n_{i}}\bm{\lambda}^{T}\bm{U}^{T}\bm{\Omega}\bm{U}\bm{\lambda}\\
 & \geq\frac{1}{\max_{i}n_{i}}C_{1}\left|\mathcal{G}_{\min}\right|\left\Vert \bm{\lambda}\right\Vert ^{2},
\end{align*}
where the last inequality is by condition (C2).
So,
\begin{align}
\frac{\max_{i}\bm{\lambda}^{T}\bm{U}_{i}^{T}\bm{\Omega}_{i}\bm{\Omega}_{i}\bm{U}_{i}\bm{\lambda}}{\bm{\lambda}^{T}\left(\sum_{i=1}^{n}\bm{U}_{i}^{T}\bm{\Omega}_{i}\bm{\Omega}_{i}\bm{U}_{i}\right)\bm{\lambda}} & \leq\left(\max_{i}n_{i}\right)\left(\max_{i}\frac{1}{n_{i}}\right)M_{1}^{2}C_{1}^{-1}\left|\mathcal{G}_{\min}\right|^{-1}\left(q+Kp\right) \nonumber \\
 & = M_{1}^{2}C_{1}^{-1}\frac{\max_{i}n_{i}}{\min_{i}n_{i}}\left|\mathcal{G}_{\min}\right|^{-1}\left(q+Kp\right)\rightarrow0, \label{eq_alimit0}
\end{align}
by assumption.

By \eqref{eq_alimit0}, we have that $\max_{i}a_{i}^{2}/\sum_{i=1}^{n}a_{i}^{2}\rightarrow0$, so \eqref{eq_clt_est} exists.

\subsection{Proof of Theorem \ref{subgroup_them2}}
In this section, we prove Theorem \ref{subgroup_them2}. As in \cite{ma2016estimating} and \cite{ma2017concave}, we define $T:\,\mathcal{M}_{\mathcal{G}}\rightarrow\mathbb{R}^{Kp}$ to be the mapping that $T\left(\bm{\beta}\right) = \bm{\alpha}$ and $T^{*}:\,\mathbb{R}^{np}\rightarrow\mathbb{R}^{Kp}$ to be the mapping that $T^{*}\left(\bm{\beta}\right)=\left( \left|\mathcal{G}_{k}\right|^{-1}\sum_{i\in\mathcal{G}_{k}}\bm{\beta}_{i}^{T},\,k=1,\dots,K\right) ^{T}$. 


Consider the following neighborhood of $\left(\bm{\eta}^{0},\bm{\beta}^{0}\right)$,
\[
\Theta=\left\{ \bm{\eta}\in\mathbb{R}^{q},\bm{\beta}\in\mathbb{R}^{np}:\,\left\Vert \bm{\eta}-\bm{\eta}^{0}\right\Vert \leq\phi_{n},\sup_{i}\left\Vert \bm{\beta}_{i}-\bm{\beta}_{i}^{0}\right\Vert \leq\phi_{n}\right\} .
\]
According to Theorem \ref{subgroup_them1}, there exists an event $E_1$ where 
$\left\Vert \bm{\eta}-\bm{\eta}^{0}\right\Vert \leq\phi_{n}$ and $\sup_{i}\left\Vert \bm{\beta}_{i}-\bm{\beta}_{i}^{0}\right\Vert \leq\phi_{n}$ such that $P\left(E_{1}\right) \geq 1 -  2\left(q+Kp\right)n^{-1}$.

Recall that the objective function to minimize is given in \eqref{eq_obj}, which has the following form
\begin{equation}
Q_{n}\left(\bm{\eta},\bm{\beta} ; \lambda, \psi\right)=\frac{1}{2}\sum_{i=1}^{n}\frac{1}{n_{i}}\sum_{h=1}^{n_{i}}\left(y_{ih}-\bm{z}_{ih}^{T}\bm{\eta}-\bm{x}_{ih}^{T}\bm{\beta}_{i}\right)^{2}+\sum_{1\leq i<j\leq n}p_{\gamma}\left(\left\Vert \bm{\beta}_{i}-\bm{\beta}_{j}\right\Vert ,c_{ij}\lambda\right).
\end{equation}
Here we show that $\left(\text{\ensuremath{(\hat{\bm{\eta}}}}^{or})^T,(\hat{\bm{\beta}}^{or})^T\right)^T$ is a strict local minimizer of the above objective function with probability approaching 1 by two steps as in \cite{ma2016estimating}.  
The first step is to show that in event $E_1$, $Q_n(\bm{\eta},\bm{\beta}^*) > Q_n(\hat{\bm{\eta}}^{or}, \hat{\bm{\beta}}^{or})$ for any $(\bm{\eta}^T,\bm{\beta}^T)^T \in \Theta$ and $(\bm{\eta}^T,\bm{\beta}^{*T})^T \neq ((\hat{\bm{\eta}}^{or})^T, (\hat{\bm{\beta}}^{or})^T)^T$, where $\bm{\beta}^{*}=T^{-1}\left(T^{*}\left(\bm{\beta}\right)\right)$ and $\bm{\beta} \in \mathbb{R}^{np}$. The proof of this step is almost the same as the first step in \cite{ma2016estimating}  
, which is omitted here.

Here we show the second step, that is, there exists an event $E_{2}$ such that $P\left(E_{2}\right)\geq 1 - 2n^{-1}$.
In the event $E_{1}\cap E_{2}$, there is a neighborhood $\Theta_{n}$ of $\left(\left(\text{\ensuremath{\hat{\bm{\eta}}}}^{or}\right)^{T},\left(\hat{\bm{\beta}}^{or}\right)^{T}\right)^{T}$, such that $Q_{n}\left(\bm{\eta},\bm{\beta}\right)\geq Q_{n}\left(\bm{\eta},\bm{\beta}^{*}\right)$ for any $\left(\bm{\eta}^{T},\bm{\beta}^{T}\right)^{T}\in\Theta_{n}\cap\Theta$ for sufficiently large $n$.

Let $\Theta_{n}=\left\{ \bm{\beta}_{i}:\sup_{i}\left\Vert \bm{\beta}_{i}-\hat{\bm{\beta}}_{i}^{or}\right\Vert \leq t_{n}\right\}$, where $t_{n}$ is a positive sequence with $t_n = o(1)$. By Taylor's expansion, for $\left(\bm{\eta}^{T},\bm{\beta}^{T}\right)^{T}\in\Theta_{n}\cap\Theta$, 
\begin{equation}
\label{eq_difference}
Q_{n}\left(\bm{\eta},\bm{\beta}\right)-Q_{n}\left(\bm{\eta},\bm{\beta}^{*}\right)=\Gamma_{1}+\Gamma_{2},
\end{equation}
where 
\begin{align*}
\Gamma_{1} & =-\left(\bm{y}-\bm{Z}\bm{\eta}-\bm{X}\bm{\beta}^{m}\right)^{T}\bm{\Omega}\bm{X}\left(\bm{\beta}-\bm{\beta}^{*}\right),\\
\Gamma_{2} & =\sum_{i=1}^{n}\frac{\partial\left[ \lambda\sum_{l<j}c_{lj}\rho_{\gamma}\left(\left\Vert \bm{\beta}_{l}^m-\bm{\beta}_{j}^m\right\Vert \right)\right]}{\partial\bm{\beta}_{i}^{T}}\left(\bm{\beta}_{i}-\bm{\beta}_{i}^{*}\right),
\end{align*}
with $\bm{\beta}^{m}=\alpha\bm{\beta}+\left(1-\alpha\right)\bm{\beta}^{*}$ for some constant $\alpha\in\left(0,1\right)$.

We have $\Gamma_{2}$ as follows,
\[
\Gamma_{2}=\lambda\sum_{i<j}c_{ij}\rho_{\gamma}^{\prime}\left(\left\Vert \bm{\beta}_{i}^{m}-\bm{\beta}_{j}^{m}\right\Vert \right)\left\Vert \bm{\beta}_{i}^{m}-\bm{\beta}_{j}^{m}\right\Vert ^{-1}\left(\bm{\beta}_{i}^{m}-\bm{\beta}_{j}^{m}\right)^{T}\left\{ \left(\bm{\beta}_{i}-\bm{\beta}_{i}^{*}\right)-\left(\bm{\beta}_{j}-\bm{\beta}_{j}^{*}\right)\right\}.
\]
For $i,j\in\mathcal{G}_{k}$, $\bm{\beta}_{i}^{*}=\bm{\beta}_{j}^{*}$ and $\bm{\beta}_{i}^{m}-\bm{\beta}_{j}^{m}=\alpha\left(\bm{\beta}_{i}-\bm{\beta}_{j}\right)$, then

\begin{align*}
\Gamma_{2} & =\lambda\sum_{k=1}^{K}\sum_{\left\{ i,j\in\mathcal{G}_{k},i<j\right\} }c_{ij}\rho_{\gamma}^{\prime}\left(\left\Vert \bm{\beta}_{i}^{m}-\bm{\beta}_{j}^{m}\right\Vert \right)\left\Vert \bm{\beta}_{i}^{m}-\bm{\beta}_{j}^{m}\right\Vert ^{-1}\left(\bm{\beta}_{i}^{m}-\bm{\beta}_{j}^{m}\right)^{T}\left(\bm{\beta}_{i}-\bm{\beta}_{j}\right)\\
 & +\lambda\sum_{k=1}^{K}\sum_{\left\{ i\in\mathcal{G}_{k},j\in\mathcal{G}_{k^{\prime}}\right\} }c_{ij}\rho_{\gamma}^{\prime}\left(\left\Vert \bm{\beta}_{i}^{m}-\bm{\beta}_{j}^{m}\right\Vert \right)\left\Vert \bm{\beta}_{i}^{m}-\bm{\beta}_{j}^{m}\right\Vert ^{-1}\left(\bm{\beta}_{i}^{m}-\bm{\beta}_{j}^{m}\right)^{T}\left\{ \left(\bm{\beta}_{i}-\bm{\beta}_{i}^{*}\right)-\left(\bm{\beta}_{j}-\bm{\beta}_{j}^{*}\right)\right\} .
\end{align*}
Since $\sup_{i}\left\Vert \bm{\beta}_{i}^{m}-\bm{\beta}_{i}^{0}\right\Vert \leq\phi_{n}$, for $k\neq k^{\prime}$, $i\in\mathcal{G}_{k}$,$j\in\mathcal{G}_{k^{\prime}}$,
\[
\left\Vert \bm{\beta}_{i}^{m}-\bm{\beta}_{j}^{m}\right\Vert \geq\min_{i\in\mathcal{G}_{k},j\in\mathcal{G}_{k^{\prime}}}\left\Vert \bm{\beta}_{i}^{0}-\bm{\beta}_{j}^{0}\right\Vert -2\max_{i}\left\Vert \bm{\beta}_{i}^{m}-\bm{\beta}_{i}^{0}\right\Vert \geq b_{n}-2\phi_{n}>a\lambda.
\]
Thus, $\rho_{\gamma}^{\prime} (\left\Vert \bm{\beta}_{i}^{m}-\bm{\beta}_{j}^{m}\right\Vert) =0$ by assumption (C1). Therefore,
\begin{equation}
\label{eq_gamma2bound}
\Gamma_{2}=\lambda\sum_{i=1}^{K}\sum_{\left\{ i,j\in\mathcal{G}_{k},i<j\right\} }c_{ij}\rho_{\gamma}^{\prime}\left(\left\Vert \bm{\beta}_{i}^{m}-\bm{\beta}_{j}^{m}\right\Vert \right)\left\Vert \bm{\beta}_{i}-\bm{\beta}_{j}\right\Vert.
\end{equation}
Also,  for $i,j \in \mathcal{G}_k$, $\sup_{i}\left\Vert \bm{\beta}_{i}^{m}-\bm{\beta}_{j}^{m}\right\Vert \leq4t_{n}$, so
$\rho_{\gamma}^{\prime}\left(\left\Vert \bm{\beta}_{i}^{m}-\bm{\beta}_{j}^{m}\right\Vert \right)\geq\rho^{\prime}\left(4t_{n}\right)$ by assumption (C1). Thus, we have
\[
\Gamma_{2}\geq\sum_{k=1}^{K}\sum_{\left\{ i,j\in\mathcal{G}_{k},i<j\right\} }\lambda c_{ij}\rho_{\gamma}^{\prime}\left(4t_{n}\right)\left\Vert \bm{\beta}_{i}-\bm{\beta}_{j}\right\Vert .
\]

Let $\bm{Q}=\left(\bm{Q}_{1}^{T},\dots,\bm{Q}_{n}^{T}\right)^{T}=\left[\left(\bm{y}-\bm{Z}\bm{\eta}-\bm{X}\bm{\beta}^{m}\right)^{T}\bm{\Omega}\bm{X}\right]^{T}$ with 
\[
\bm{Q}_{i}=\frac{1}{n_{i}}\sum_{h=1}^{n_{i}}\left(y_{ih}-\bm{z}_{ih}^{T}\bm{\eta}-\bm{x}_{ih}^{T}\bm{\beta}_{i}^m\right)\bm{x}_{ih}.
\]
We have,
\begin{align}
\label{eq_gamma1bound}
\Gamma_{1} & =-\left(\bm{y}-\bm{Z}\bm{\eta}-\bm{X}\bm{\beta}^{m}\right)^{T}\bm{\Omega}\bm{X}\left(\bm{\beta}-\bm{\beta}^{*}\right) \nonumber \\
 & =-\bm{Q}^{T}\left(\bm{\beta}-\bm{\beta}^{*}\right)\nonumber \\
 & =-\sum_{k=1}^{K}\sum_{\left\{ i,j\in\mathcal{G}_{k},i<j\right\} }\frac{\left(\bm{Q}_{i}-\bm{Q}_{j}\right)^{T}\left(\bm{\beta}_{i}-\bm{\beta}_{j}\right)}{\left|\mathcal{G}_{k}\right|}.
\end{align}
Moreover, 
\[
\bm{Q}_{i}=\frac{1}{n_{i}}\sum_{h=1}^{n_{i}}\left(\epsilon_{ih}+\bm{z}_{ih}^{T}\left(\bm{\eta}^{0}-\bm{\eta}\right)+\bm{x}_{ih}^{T}\left(\bm{\beta}_{i}^{0}-\bm{\beta}_{i}^{m}\right)\right)\bm{x}_{ih},
\]
so 
\begin{align*}
\sup_{i}\left\Vert \bm{Q}_{i}\right\Vert  & \leq\sup_{i,h}\left\Vert \bm{x}_{ih}\right\Vert \left(\left\Vert \bm{\xi}\right\Vert _{\infty}+\sup_{i,h}\left\Vert \bm{z}_{ih}\right\Vert \left\Vert \bm{\eta}^{0}-\bm{\eta}\right\Vert +\sup_{i,h}\left\Vert \bm{x}_{ih}\right\Vert \left\Vert \bm{\beta}_{i}^{0}-\bm{\beta}_{i}^{m}\right\Vert \right)\\
 & \leq C_{2}\sqrt{p}\left(\left\Vert \bm{\xi}\right\Vert _{\infty}+C_{3}\sqrt{q}\phi_{n}+C_{2}\sqrt{p}\phi_{n}\right),
\end{align*}
where $\bm{\xi}=\left(\xi_{1,}\dots,\xi_{n}\right)^{T}$ with $\xi_{i}=\frac{1}{n_{i}}\sum_{h=1}^{n}\epsilon_{ih}$. According to Condition (C3), 
\begin{align*}
P\left(\left\Vert \bm{\xi}\right\Vert _{\infty}>\sqrt{2c_{1}^{-1}}\sqrt{\log n/\min n_{i}}\right) & \leq\sum_{i=1}^{n}P\left(\left|\xi_{i}\right|>\sqrt{2c_{1}^{-1}}\sqrt{\log n/\min n_{i}}\right)\\
 & =\sum_{i=1}^{n}P\left(\left|\frac{1}{n_{i}}\sum_{j=1}^{n_{i}}\epsilon_{ij}\right|>\sqrt{2c_{1}^{-1}}\sqrt{\log n/\min n_{i}}\right)\\
 & \leq\sum_{i=1}^{n}P\left(\left|\frac{1}{n_{i}}\sum_{j=1}^{n_{i}}\epsilon_{ij}\right|>\sqrt{2c_{1}^{-1}}\sqrt{\log n/n_{i}}\right)\\
 & \leq2\sum_{i=1}^{n}\exp\left\{ -c_{1}2c_{1}^{-1}\log n\right\} \leq\frac{2}{n}.
\end{align*}

Thus, there exists an event $E_{2}$ such that $P\left(E_{2}\right)\geq 1 - 2n^{-1}$ and 
\[
\sup_{i}\left\Vert \bm{Q}_{i}\right\Vert \leq C_{2}\sqrt{p}\left(\sqrt{2c_{1}^{-1}}\sqrt{\log n/\min_{i}n_{i}}+C_{3}\sqrt{q}\phi_{n}+C_{2}\sqrt{p}\phi_{n}\right).
\]
Thus,
\begin{align}
 & \left|\frac{\left(\bm{Q}_{i}-\bm{Q}_{j}\right)^{T}\left(\bm{\beta}_{i}-\bm{\beta}_{j}\right)}{\left|\mathcal{G}_{k}\right|}\right| \nonumber \\
\leq & 2\left|\mathcal{G}_{\min}\right|^{-1}\sup_{i}\left\Vert \bm{Q}_{i}\right\Vert \left\Vert \bm{\beta}_{i}-\bm{\beta}_{j}\right\Vert \nonumber \\
\leq & 2C_{2}\left|\mathcal{G}_{\min}\right|^{-1}\sqrt{p}\left(\sqrt{2c_{1}^{-1}}\sqrt{\log n/\min_{i}n_{i}}+C_{3}\sqrt{q}\phi_{n}+C_{2}\sqrt{p}\phi_{n}\right)\left\Vert \bm{\beta}_{i}-\bm{\beta}_{j}\right\Vert. \label{eq_supubound}
\end{align}

Combining \eqref{eq_gamma2bound},  \eqref{eq_gamma1bound} and \eqref{eq_supubound},  \eqref{eq_difference}  follows that 
{\small
\begin{align*}
 & Q_{n}\left(\bm{\eta},\bm{\beta}\right)-Q_{n}\left(\bm{\eta},\bm{\beta}^{*}\right)\\
\geq & \sum_{k=1}^{K}\sum_{\left\{ i,j\in\mathcal{G}_{k},i<j\right\} }\left\{ \lambda c_{ij}\rho^{\prime}\left(4t_{n}\right)-2C_{2}\left|\mathcal{G}_{\min}\right|^{-1}\sqrt{p}\left(\sqrt{2c_{1}^{-1}}\sqrt{\frac{\log n}{\min_{i}n_{i}}}+C_{3}\sqrt{q}\phi_{n}+C_{2}\sqrt{p}\phi_{n}\right)\right\} \left\Vert \bm{\beta}_{i}-\bm{\beta}_{j}\right\Vert .
\end{align*}
}
As $t_{n}=o\left(1\right)$, $\rho^{\prime}\left(4t_{n}\right)\rightarrow1$. Since $\left|\mathcal{G}_{\text{min}}\right|\gg  \left(q+Kp\right)^{1/2} \max \left(\sqrt{\frac{n}{\min_{i}n_{i}}\log n},\left(q+Kp\right)^{1/2}\right)$, $p = o(n)$ and $q = o(n)$,  then $\left|\mathcal{G}_{\min}\right|^{-1}p =o\left(1\right)$ and $\left|\mathcal{G}_{\min}\right|^{-1}\sqrt{pq} =o\left(1\right)$. Thus, $\lambda \gg\left|\mathcal{G}_{\min}\right|^{-1}\sqrt{p}\sqrt{\frac{\log n}{\min n_{i}}}$, $\lambda\gg\left|\mathcal{G}_{\min}\right|^{-1}\sqrt{pq}\phi_n$ and $\lambda\gg\left|\mathcal{G}_{\min}\right|^{-1}p\phi_{n}$. Therefore, $Q_{n}\left(\bm{\eta},\bm{\beta}\right)-Q_{n}\left(\bm{\eta},\bm{\beta}^{*}\right)\geq0$ for sufficiently large $n$ by the assumption  (C4) that $c_{ij}$'s are bounded if $i$ and $j$ are in the same group.

Therefore, combining the two steps, we will have that $Q_{n}\left(\bm{\eta},\bm{\beta}\right) > Q_n( \hat{\bm{\eta}}^{or}, \hat{\bm{\beta}}^{or})$ for any $\left(\bm{\eta}^{T},\bm{\beta}^{T}\right)^{T}\in\Theta_{n}\cap\Theta$ and $(\bm{\eta}^T,\bm{\beta}^{T})^T \neq ((\hat{\bm{\eta}}^{or})^T, (\hat{\bm{\beta}}^{or})^T)^T$. This shows that $((\hat{\bm{\eta}}^{or})^T, (\hat{\bm{\beta}}^{or})^T)^T$ is a strict local minimizer of the objective function \eqref{eq_obj} on $E_1 \cap E_2$ with probability at least $ 1- 2(K+p+1)n^{-1}$ for sufficiently large $n$.

\end{appendices}
\bibliographystyle{apalike} 
\bibliography{ref_subgroups.bib}

\end{document}